\documentclass[a4paper, USenglish, cleveref, autoref]{lipics-v2021}
\pdfoutput=1 
\usepackage{amsmath, amssymb, amsthm}
\usepackage{graphicx}
\usepackage{hyperref}
\usepackage{float}
\usepackage{tikz}
\usepackage{xcolor}

\hideLIPIcs
\nolinenumbers

\title{NP-Completeness Proofs of Puzzles using the T-Metacell Framework}
\titlerunning{NP-Completeness Proofs of Puzzles}

\author{Nattapol Kiatchaipipat}{University of Wisconsin-Madison, USA}{nkiatchaipip@wisc.edu}{}{}
\author{Suthee Ruangwises}{Chulalongkorn University, Thailand}{suthee@cp.eng.chula.ac.th}{https://orcid.org/0000-0002-2820-1301}{}

\authorrunning{N. Kiatchaipipat and S. Ruangwises}
\Copyright{Nattapol Kiatchaipipat and Suthee Ruangwises}



\begin{document}

\maketitle

\begin{abstract}
Pencil puzzles are puzzles that can be solved by writing down solutions on a paper, using only logical reasoning. In this paper, we utilize the "T-metacell" framework developed by Tang \cite{tcell} and the MIT Hardness Group \cite{37puzzles} to prove the NP-completeness of four new pencil puzzles: Grand Tour, Entry Exit, Zahlenschlange, and Yagit. Additionally, the first three are also proven to be ASP-complete. The results demonstrate how versatile the framework is, offering new insights into the computational complexity of problems with various constraints.
\end{abstract}

\section{Introduction}
Pencil puzzles are puzzles that can be solved by writing down solutions on a paper, using only logical reasoning. Examples of pencil puzzles include Sudoku, Makaro, and Slitherlink.

The computational complexity of pencil puzzles has attracted considerable interest for their rich connection to constraint satisfaction and graph theory. Many popular pencil puzzles have been proved to be NP-complete using various techniques, including Five Cells \cite{fivecells}, Goishi Hiroi \cite{goishi}, Hashiwokakero \cite{bridges}, Herugolf \cite{makaro}, Heyawake \cite{heyawake}, Kakuro \cite{sudoku}, Makaro \cite{makaro}, Numberlink \cite{numberlink}, Slitherlink \cite{sudoku}, Sudoku \cite{sudoku}, Sumplete \cite{sumplete}, Tatamibari \cite{tatamibari}, Tilepaint \cite{fivecells}, and Zeiger \cite{zeiger}.

\subsection{ASP-Completeness and T-Metacell Framework}
An NP search problem is ASP-complete if for each NP search problem, there exist a polynomial time reduction to it along with a polynomial time bijection between the solutions of both problems. The decision version of a ASP-complete problem is NP-complete, the counting version is \#P-complete, and the $k$-ASP P problem — given an instance of $P$ and $k$ solutions, find another solution — is NP-complete for any $k \geq 0$ \cite{sudoku}.

In 2022, Tang \cite{tcell} proposed a ``T-metacell" framework which uses grid graph Hamiltonicity to prove NP-hardness of puzzles. A T-metacell is any gadget that represents a vertex of degree 3 in a grid graph. Each gadget must have exactly three exits out of the four sides, and must be reflectable and rotatable, and a path can be made between two T-metacells only when they have their exits aligned. Solving Hamiltonian cycle on a grid of T-metacells was known to be NP-complete \cite{hamiltonian}.

Tang's framework was recently used by Deurloo et al. \cite{5puzzles} to prove the NP-completeness of five puzzles, and by the MIT Hardness Group \cite{37puzzles} to prove the NP-completeness of 37 puzzles. The MIT hardness group also proved that if the solution for each pair of exits in a T-metacell is unique, then solving the problem as a whole can be further classified as ASP-complete. Additionally, they extended the framework to allow for T-metacell with one forced exit. This, however, requires the cell to be ``asymmetric", where the two unforced exits must be adjacent to each other. The results used in this work are as follow.

\begin{enumerate}
    \item Finding Hamiltonian cycles on a rectangular grid of undirected T-metacells is ASP-complete \cite[Corollary 5.1]{37puzzles}.
    \item Finding Hamiltonian cycles on a rectangular grid of asymmetric forced-edge undirected T-metacells is ASP-complete \cite[Corollary 5.3]{37puzzles}.
\end{enumerate}

All puzzles discussed in this paper are restricted to traversal of only orthogonally adjacent cells.

\subsection{Our Contribution}
Building on the methods of \cite{37puzzles}, which developed the existing T-metacell framework to be sufficient proof of ASP-completeness, we create a series of reductions that establish the ASP-completeness of three new diverse puzzles: Grand Tour, Entry Exit, and Zahlenschlange. Additionally, we used the non-unique version of the framework to prove the NP-completeness of another puzzle: Yagit.

\section{Hardness Proofs}
\subsection{Grand Tour}
In Grand Tour, the goal is to draw a single loop that goes through every vertex in the graph exactly once, with the added rule that there can be edges that must be used. 

The rule for this question is fairly general and may be useful for proving ASP-completeness of future problems related to Hamiltonian cycles.

\begin{figure}[h]
\centering

\begin{subfigure}{0.45\textwidth}
\centering
\begin{tikzpicture}[scale=0.7]
    \draw[step=1cm,gray,very thin] (0,0) grid (5,5);
    \draw[line width=1pt] (0,0) rectangle (5,5);

    \draw[red, ultra thick]
        (3,3) -- (3,2)         
        (4,3) -- (4,2) 
        (3,1) -- (3,0);         
\end{tikzpicture}
\caption{Unfilled Example}
\end{subfigure}
\hfill
\begin{subfigure}{0.45\textwidth}
\centering
\begin{tikzpicture}[scale=0.7]
    \draw[step=1cm,gray,very thin] (0,0) grid (5,5);
    \draw[line width=1pt] (0,0) rectangle (5,5);

    \draw[red, ultra thick]
        (3,3) -- (3,2)         
        (4,3) -- (4,2) 
        (3,1) -- (3,0);         
    \draw[blue, ultra thick]
        (3,0) -- (2,0) -- (2,4) -- (1,4) -- (1,0) -- (0,0) -- (0,5) -- (5,5) -- (5,0) -- (4,0) -- (4,2)
        (3,1) -- (3,2)
        (3,3) -- (3,4) -- (4,4) -- (4,3);   
\end{tikzpicture}
\caption{Filled Example}
\end{subfigure}
\label{1}
\end{figure}

More examples can be found at \cite{grandtour2025}.

\begin{figure}[H]
\centering

\begin{subfigure}[t]{0.45\textwidth}
\centering
\begin{tikzpicture}[scale=0.8]
    \draw[step=1cm,gray,very thin] (0,0) grid (6,6);
    \draw[line width=1pt] (0,0) rectangle (6,6);

    \draw[red, ultra thick]
        (1,3) -- (1,5) -- (5,5) -- (5,4) -- (4,4)
        (4,2) -- (4,1) -- (5,1) -- (5,3)
        (3,1) -- (1,1) -- (1,2) -- (2,2);
\end{tikzpicture}
\caption{5×5 undirected T-metacell}
\end{subfigure}
\hfill
\begin{subfigure}[t]{0.45\textwidth}
\centering
\begin{tikzpicture}[scale=0.8]
    \draw[step=1cm,gray,very thin] (0,0) grid (6,6);
    \draw[line width=1pt] (0,0) rectangle (6,6);

    \draw[red, ultra thick]
        (1,3) -- (1,5) -- (5,5) -- (5,4) -- (4,4)
        (4,2) -- (4,1) -- (5,1) -- (5,3)
        (3,1) -- (1,1) -- (1,2) -- (2,2);
    \draw[blue, ultra thick]
        (0,3) -- (1,3)
        (2,2) -- (2,4) -- (4,4)
        (3,1) -- (3,3) -- (4,3) -- (4,2)
        (5,3) -- (6,3);
\end{tikzpicture}
\caption{5×5 filled undirected T-metacell}
\end{subfigure}

\vspace{1em}

\begin{subfigure}[t]{0.45\textwidth}
\centering
\begin{tikzpicture}[scale=0.8]
    \draw[step=1cm,gray,very thin] (0,0) grid (6,6);
    \draw[line width=1pt] (0,0) rectangle (6,6);

    \draw[red, ultra thick]
        (1,3) -- (1,5) -- (5,5) -- (5,4) -- (4,4)
        (4,2) -- (4,1) -- (5,1) -- (5,3)
        (3,1) -- (1,1) -- (1,2) -- (2,2);
    \draw[blue, ultra thick]
        (0,3) -- (1,3)
        (5,3) -- (2,3) -- (2,4) -- (4,4)
        (2,2) -- (4,2)
        (3,0) -- (3,1);
\end{tikzpicture}
\caption{5×5 filled undirected T-metacell}
\end{subfigure}
\hfill
\begin{subfigure}[t]{0.45\textwidth}
\centering
\begin{tikzpicture}[scale=0.8]
    \draw[step=1cm,gray,very thin] (0,0) grid (6,6);
    \draw[line width=1pt] (0,0) rectangle (6,6);

    \draw[red, ultra thick]
        (1,3) -- (1,5) -- (5,5) -- (5,4) -- (4,4)
        (4,2) -- (4,1) -- (5,1) -- (5,3)
        (3,1) -- (1,1) -- (1,2) -- (2,2);
    \draw[blue, ultra thick]
        (5,3) -- (6,3)
        (1,3) -- (2,3) -- (2,4) -- (3,4) -- (3,2) -- (2,2)
        (4,4) -- (4,2)
        (3,0) -- (3,1);
\end{tikzpicture}
\caption{5×5 filled undirected T-metacell}
\end{subfigure}

\caption{5×5 undirected T-metacell for Grand Tour with all exit choices}
\label{2}
\end{figure}

\cref{2} contains a 5x5 undirected T-metacell for Grand Tour. All possible exit choices can be done as shown. This works simply by the fact that there are three exits. Since the rule requires the loop to visit every vertex exactly once, and the forced edges require the loop to use at least two exits, all exit points must be used as the loop passes through the cell.

In fact, we can generalize this to a class of structure that all functions as a T-metacell for Grand Tour. For any grid of size $4n + 1$ for $n \in \mathbb{N}$ we can construct the same shape with the exits aligned to the middle of each side as shown in \cref{3}. The pattern for each exit choices is one of possibly many ways to trace through the cell. Due to the uniqueness requirement however, this design is sufficient for proving NP-completeness but not ASP-completeness.

It is clear that as long as the grid expands by 4, the cell can be traversed in a similar pattern for any pair of exits. As the number of unforced edges grows faster than the number of forced edges, Grand Tour is NP-complete for arbitrary small proportion of forced edges to unforced edges.

\begin{figure}[H]
\centering

\begin{subfigure}[t]{0.45\textwidth}
\centering
\begin{tikzpicture}[scale=0.4]
    \draw[step=1cm,gray,very thin] (0,0) grid (10,10);
    \draw[line width=1pt] (0,0) rectangle (10,10);

    \draw[red, ultra thick]
        (2,4) -- (1,4) -- (1,1) -- (5,1)
        (6,2) -- (6,1) -- (9,1) -- (9,5)
        (1,5) -- (1,9) -- (9,9) -- (9,6) -- (8,6);

\end{tikzpicture}
\caption{9x9 undirected T-metacell}
\end{subfigure}
\hfill
\begin{subfigure}[t]{0.45\textwidth}
\centering
\begin{tikzpicture}[scale=0.4]
    \draw[step=1cm,gray,very thin] (0,0) grid (10,10);
    \draw[line width=1pt] (0,0) rectangle (10,10);

    \draw[red, ultra thick]
        (2,4) -- (1,4) -- (1,1) -- (5,1)
        (6,2) -- (6,1) -- (9,1) -- (9,5)
        (1,5) -- (1,9) -- (9,9) -- (9,6) -- (8,6);
    \draw[blue, ultra thick]
        (0,5) -- (1,5)
        (8,6) -- (3,6) -- (3,7) -- (8,7) -- (8,8) -- (2,8) -- (2,4)
        (5,1) -- ( 5,2) -- ( 2,2) -- (2,3) -- (5,3) -- (5,4) -- (3, 4) -- (3,5) -- ( 8,5) -- (8,2) -- (7,2) -- (7,4) -- (6,4) -- ( 6,2)
        (10,5) -- (9,5);

\end{tikzpicture}
\caption{9x9 filled undirected T-metacell}
\end{subfigure}

\vspace{0.1em}

\begin{subfigure}[t]{0.45\textwidth}
\centering
\begin{tikzpicture}[scale=0.4]
    \draw[step=1cm,gray,very thin] (0,0) grid (10,10);
    \draw[line width=1pt] (0,0) rectangle (10,10);

    \draw[red, ultra thick]
        (2,4) -- (1,4) -- (1,1) -- (5,1)
        (6,2) -- (6,1) -- (9,1) -- (9,5)
        (1,5) -- (1,9) -- (9,9) -- (9,6) -- (8,6);
    \draw[blue, ultra thick]
        (0,5) -- (1,5)
        (2,4) -- (2,2) -- (3,2) -- (3,4) -- (4,4) -- (4,2) -- (5,2) -- (5,4) -- (6,4) -- (6,3) -- (7,3) -- (7,4) -- (8,4) -- (8,2) -- (6,2)
        (9,5) -- (2,5) -- (2,8) -- (8,8) -- (8,7) -- (3,7) -- (3,6) -- (8,6)
        (5,1) -- (5,0);

\end{tikzpicture}
\caption{9x9 filled undirected T-metacell}
\end{subfigure}
\hfill
\begin{subfigure}[t]{0.45\textwidth}
\centering
\begin{tikzpicture}[scale=0.4]
    \draw[step=1cm,gray,very thin] (0,0) grid (10,10);
    \draw[line width=1pt] (0,0) rectangle (10,10);

    \draw[red, ultra thick]
        (2,4) -- (1,4) -- (1,1) -- (5,1)
        (6,2) -- (6,1) -- (9,1) -- (9,5)
        (1,5) -- (1,9) -- (9,9) -- (9,6) -- (8,6);
    \draw[blue, ultra thick]
        (10,5) -- (9,5)
        (1,5) -- (2,5) -- (2,8) -- (5,8) -- (5,2) -- (4,2) -- (4,7) -- (3,7) -- (3,2) -- (2,2) -- (2,4)
        (6,2) -- (8,2) -- (8,3) -- ( 6,3) -- (6,4) -- (8,4) -- (8,5) -- (6,5) -- (6,6) -- ( 7,6) -- (7,7) -- (6,7) -- (6,8) -- (7,8) -- (8,8) -- (8,6)
        (5,1) -- (5,0);

\end{tikzpicture}
\caption{9x9 filled undirected T-metacell}
\end{subfigure}

\vspace{0.2em}

\begin{subfigure}[t]{0.45\textwidth}
\centering
\begin{tikzpicture}[scale=0.3]
    \draw[step=1cm,gray,very thin] (0,0) grid (18,18);
    \draw[line width=1pt] (0,0) rectangle (18,18);

    \draw[red, ultra thick]
        (2,8) -- (1,8) -- (1,1) -- (9,1)
        (1,9) -- (1,17) -- (17,17) -- (17,10) -- ( 16,10)
        (17,9) -- (17,1) -- (10,1) -- (10,2);
        
\end{tikzpicture}

\caption{17x17 filled undirected T-metacell}
\end{subfigure}
\hfill
\begin{subfigure}[t]{0.45\textwidth}
\centering
\begin{tikzpicture}[scale=0.3]
    \draw[step=1cm,gray,very thin] (0,0) grid (18,18);
    \draw[line width=1pt] (0,0) rectangle (18,18);

    \draw[red, ultra thick]
        (2,8) -- (1,8) -- (1,1) -- (9,1)
        (1,9) -- (1,17) -- (17,17) -- (17,10) -- ( 16,10)
        (17,9) -- (17,1) -- (10,1) -- (10,2);
    \draw[blue, ultra thick]
        (0,9) -- (1,9)
        (2,8) -- ( 2,16) -- (16,16) -- (16,15) -- (3,15) -- (3,14) -- (16,14) -- ( 16,13) -- (3,13) -- (3,12) -- (16,12) -- (16,11) -- (3,11) -- (3,10) -- ( 16,10)
        (9,1) -- (9,2) -- (2,2) -- (2,3) -- (9,3) -- (9,4) -- ( 2,4) -- (2,5) -- (9,5) -- ( 9,6) -- (2,6) -- (2,7) -- (9,7) -- (9,8) -- (3,8) -- (3,9) -- (16,9) -- (16,2) -- (15,2) -- (15,8) -- (14,8) -- (14,2) -- (13,2) -- (13,8) -- (12,8) -- (12,2) -- (11,2) -- (11,8) -- (10,8) -- (10,2)
        (17,9) -- (18,9);
        
\end{tikzpicture}
\caption{17x17 filled undirected T-metacell}
\end{subfigure}

\vspace{0.2em}

\begin{subfigure}[t]{0.45\textwidth}
\centering
\begin{tikzpicture}[scale=0.3]
    \draw[step=1cm,gray,very thin] (0,0) grid (18,18);
    \draw[line width=1pt] (0,0) rectangle (18,18);

    \draw[red, ultra thick]
        (2,8) -- (1,8) -- (1,1) -- (9,1)
        (1,9) -- (1,17) -- (17,17) -- (17,10) -- ( 16,10)
        (17,9) -- (17,1) -- (10,1) -- (10,2);
    \draw[blue, ultra thick]
        (0,9) -- (1,9)
        (17,9) -- (2,9) -- ( 2,16) -- (16,16) -- (16,15) -- (3,15) -- (3,14) -- (16,14) -- ( 16,13) -- (3,13) -- (3,12) -- (16,12) -- (16,11) -- (3,11) -- (3,10) -- ( 16,10)

        (10,2) -- (16,2) -- (16,8) -- (15,8) -- (15,3) -- (14,3) -- (14,8) -- (13,8) -- (13,3) -- (12,3) -- (12,8) -- (11,8) -- (11,3) -- (10,3) -- (10,8) -- (9,8) -- (9,2) -- (8,2) -- (8,8) -- (7,8) -- (7,2) -- (6,2) -- (6,8) -- (5,8) -- (5,2) -- (4,2) -- (4,8) -- (3,8) -- (3,2) -- (2,2) -- (2,8)
        (9,1) -- (9,0);
        
\end{tikzpicture}

\caption{17x17 filled undirected T-metacell}
\end{subfigure}
\hfill
\begin{subfigure}[t]{0.45\textwidth}
\centering
\begin{tikzpicture}[scale=0.3]
    \draw[step=1cm,gray,very thin] (0,0) grid (18,18);
    \draw[line width=1pt] (0,0) rectangle (18,18);

    \draw[red, ultra thick]
        (2,8) -- (1,8) -- (1,1) -- (9,1)
        (1,9) -- (1,17) -- (17,17) -- (17,10) -- ( 16,10)
        (17,9) -- (17,1) -- (10,1) -- (10,2);
    \draw[blue, ultra thick]
        (17,9) -- (18,9)
        (1,9) -- (2,9) -- (2,16) -- (3,16) -- (9,16) -- (9,2) -- (8,2) -- (8,15) -- (7,15) -- (7,2) -- (6,2) -- (6,15) -- (5,15) -- (5,2) -- (4,2) -- (4,15) -- (3,15) -- (3,2) -- (2,2) -- (2,8)

        (10,2) -- (16,2) -- (16,3) -- (10,3) -- (10,4) -- (16,4) -- (16,5) -- (10,5) -- (10, 6) -- (16, 6) -- (16, 7) -- (10,7) -- (10,8) -- (16,8) -- (16,9) -- (10,9) -- (10,10) -- (15,10) -- (15,11) -- (10,11) -- (10,12) -- (15,12) -- (15,13) -- (10,13) -- (10, 14) -- (15, 14) -- (15,15) -- (10,15) -- (10,16) -- (16,16) -- (16, 10)
        (9,1) -- (9,0);
        
\end{tikzpicture}

\caption{17x17 filled undirected T-metacell}
\end{subfigure}

\caption{9x9 and 17x17 undirected T-metacell for Grand Tour with all exit choices}
\label{3}
\end{figure}

\cref{4} shows an asymmetric forced-edge undirected T-metacell with a unique solution for every pair of exits, hence, proving that Grand Tour is ASP-complete. Note that this construction forces the left exit to be used. 

The maximum number of forced edge in a grid graph is $2n$ which induces a Hamiltonian cycle, and cannot form a T-metacell. This design has the current maximum proportion of forced edges to unforced edges for Grand Tour to be ASP-complete, where every vertex has at least 1 forced edge. 

\begin{figure}[H]
    \centering
    \begin{tikzpicture}[scale=0.5]
    \draw[step=1cm,gray,very thin] (0,0) grid (6,6);
    \draw[line width=1pt] (0,0) rectangle (6,6);

    \draw[red, ultra thick]
        (1,3) -- (1,1) -- (2,1) -- (2,4) -- (1,4) -- (1,5) -- (5,5) -- (5,4) -- (3,4) -- (3,3)
        (3,2) -- (4,2) -- (4,3)
        (3,1) -- (5,1) -- (5,3);

    \end{tikzpicture}
    \caption{5x5 Asymmetrical Forced-edge Undirected T-metacell }
    \label{4}
\end{figure}

Notably, this is a parallel to the MIT Hardness Group's paper's problem of Hamiltonian cycle in a Max-degree-3 spanning subgraph of rectangular grid graph, where every vertex must have one forbidden edge, which was proven to be ASP-complete.

\subsection{Entry-Exit}
In Entry-Exit, the grid is divided into multiple regions, and the goal is to find a single closed loop that traverses every cell exactly once while entering and exiting each region exactly once. 

\begin{figure}[H]
    \centering
    \begin{subfigure}[t]{0.45\textwidth}
    \centering
    \begin{tikzpicture}[scale=0.5]
    \draw[step=1cm,gray,very thin] (1,1) grid (7,7);
    \draw[violet, ultra thick] (1,1) rectangle (7,7);

    \draw[violet, ultra thick]
        (1,3) -- (2,3) -- (2,2) -- (3,2) -- (3,1)
        (3,2) -- (4,2) -- (4,1)
        (4,2) -- (6,2) -- (6,1)
        (6,2) -- (6,6) -- ( 4,6) -- (4,7)
        (4,6) -- (4,5) -- (5,5) -- (5,3) -- (4,3) -- (4,2)
        (3,2) -- (3,6) -- (2,6) -- (2,7)
        (1,5) -- (2,5) -- (2,4) -- (5,4);

    \end{tikzpicture}
    \caption{Unfilled Example}
    \end{subfigure}
    \hfill
    \begin{subfigure}[t]{0.45\textwidth}
    \centering
    \begin{tikzpicture}[scale=0.5]
    \draw[step=1cm,gray,very thin] (1,1) grid (7,7);
    \draw[violet, ultra thick] (1,1) rectangle (7,7);

    \draw[violet, ultra thick]
        (1,3) -- (2,3) -- (2,2) -- (3,2) -- (3,1)
        (3,2) -- (4,2) -- (4,1)
        (4,2) -- (6,2) -- (6,1)
        (6,2) -- (6,6) -- ( 4,6) -- (4,7)
        (4,6) -- (4,5) -- (5,5) -- (5,3) -- (4,3) -- (4,2)
        (3,2) -- (3,6) -- (2,6) -- (2,7)
        (1,5) -- (2,5) -- (2,4) -- (5,4);

    \draw[blue, ultra thick]
        (1.5, 1.5) -- (6.5, 1.5) -- (6.5,6.5) -- (4.5,6.5) -- (4.5,5.5) -- (5.5,5.5) -- (5.5,2.5) -- (3.5,2.5) -- (3.5,3.5) -- (4.5,3.5) -- (4.5,4.5) -- (3.5,4.5) -- (3.5,6.5) -- (1.5,6.5) -- (1.5,5.5) -- (2.5,5.5) -- (2.5,4.5) -- (1.5,4.5) -- (1.5,3.5) -- (2.5,3.5) -- (2.5,2.5) -- (1.5,2.5) -- (1.5,1.5);

    \end{tikzpicture}
    \caption{Filled Example}
    \end{subfigure}
		\label{5}
\end{figure}

More examples can be found at \cite{Entry-Exit2025}.

This puzzle has a T-metacell that is very similar to Grand Tour. We can rely on the fact that the loop must enter each region only once, and since the width of each region is one, entering in the middle is not possible. \cref{6} shows the step-by-step reasoning behind the uniqueness of traversing the T-metacell. The internal region of the gadget can be subdivided into individual grid depending on if the variant requires each region to have width exactly 1.

Note that, similar to the T-metacell in Grand Tour, the left exit is forced since using just the other two exits is impossible. Step 6 is crucial and is implied because if the path from the left exit preemptively connects to the lower right quadrant then the rest of the cell will not be traversed.

\begin{figure}[H]
    \centering

\begin{subfigure}[t]{0.45\textwidth}
\centering
\begin{tikzpicture}[scale=0.4]
    \draw[step=1cm,gray,very thin] (0,0) grid (11,11);
    \draw[violet, ultra thick] (1,1) rectangle (10,10);

    \draw[violet, ultra thick]
        (1,6) -- (3,6) -- (3,7) -- (2,7) -- (2,9) -- (9,9) -- (9,7) -- (5,7) -- (5,6) -- (10,6)
        (2,6) -- (2,2) -- (4,2) -- (4,6) -- (5,6) -- (5,1)
        (5,5) -- (6,5) -- (6,6)
        (5,4) -- (7,4) -- (7,6)
        (5,3) -- (8,3) -- (8,6)
        (5,2) -- (9,2) -- (9,6);

\end{tikzpicture}

\caption{9x9 forced-edge undirected T-metacell}

\end{subfigure}
\hfill
\begin{subfigure}[t]{0.45\textwidth}
\centering
\begin{tikzpicture}[scale=0.4]

    \draw[step=1cm,gray,very thin] (0,0) grid (11,11);
    \draw[violet, ultra thick] (1,1) rectangle (10,10);

    \draw[violet, ultra thick]
        (1,6) -- (3,6) -- (3,7) -- (2,7) -- (2,9) -- (9,9) -- (9,7) -- (5,7) -- (5,6) -- (10,6)
        (2,6) -- (2,2) -- (4,2) -- (4,6) -- (5,6) -- (5,1)
        (5,5) -- (6,5) -- (6,6)
        (5,4) -- (7,4) -- (7,6)
        (5,3) -- (8,3) -- (8,6)
        (5,2) -- (9,2) -- (9,6);

    \draw[red, ultra thick]
        (0.5, 5.5) -- (1.5, 5.5) -- (1.5,1.5) -- (4.5,1.5) -- (4.5, 5.5)
        (2.5, 6.5) -- (1.5, 6.5) -- (1.5, 9.5) -- (9.5, 9.5) -- (9.5, 6.5) -- (5.5 , 6.5)
        (5.5, 1.5) -- (9.5,1.5) -- (9.5, 5.5)
        (5.5, 2.5) -- (8.5, 2.5) -- (8.5, 5.5)
        (5.5, 3.5) -- (7.5, 3.5) -- (7.5, 5.5)
        (5.5, 4.5) -- (6.5, 4.5) -- (6.5, 5.5);

\end{tikzpicture}
\caption{Step 1}
\end{subfigure}

\vspace{1em}

\begin{subfigure}[t]{0.45\textwidth}
\centering
\begin{tikzpicture}[scale=0.4]
    \draw[step=1cm,gray,very thin] (0,0) grid (11,11);
    \draw[violet, ultra thick] (1,1) rectangle (10,10);

    \draw[violet, ultra thick]
        (1,6) -- (3,6) -- (3,7) -- (2,7) -- (2,9) -- (9,9) -- (9,7) -- (5,7) -- (5,6) -- (10,6)
        (2,6) -- (2,2) -- (4,2) -- (4,6) -- (5,6) -- (5,1)
        (5,5) -- (6,5) -- (6,6)
        (5,4) -- (7,4) -- (7,6)
        (5,3) -- (8,3) -- (8,6)
        (5,2) -- (9,2) -- (9,6);

    \draw[blue, ultra thick]
        (0.5, 5.5) -- (1.5, 5.5) -- (1.5,1.5) -- (4.5,1.5) -- (4.5, 5.5)
        (2.5, 6.5) -- (1.5, 6.5) -- (1.5, 9.5) -- (9.5, 9.5) -- (9.5, 6.5) -- (5.5 , 6.5)
        (5.5, 1.5) -- (9.5,1.5) -- (9.5, 5.5)
        (5.5, 2.5) -- (8.5, 2.5) -- (8.5, 5.5)
        (5.5, 3.5) -- (7.5, 3.5) -- (7.5, 5.5)
        (5.5, 4.5) -- (6.5, 4.5) -- (6.5, 5.5);

    \draw[red, ultra thick]
        (2.5, 3.5) -- (2.5, 2.5) -- (3.5, 2.5) -- (3.5, 3.5)
        (7.5, 8.5) -- (8.5, 8.5) -- (8.5, 7.5) -- (7.5, 7.5)
        (3.5, 8.5) -- (2.5, 8.5) -- (2.5,7.5);

\end{tikzpicture}

\caption{Step 2}

\end{subfigure}
\hfill
\begin{subfigure}[t]{0.45\textwidth}
\centering
\begin{tikzpicture}[scale=0.4]

    \draw[step=1cm,gray,very thin] (0,0) grid (11,11);
    \draw[violet, ultra thick] (1,1) rectangle (10,10);

    \draw[violet, ultra thick]
        (1,6) -- (3,6) -- (3,7) -- (2,7) -- (2,9) -- (9,9) -- (9,7) -- (5,7) -- (5,6) -- (10,6)
        (2,6) -- (2,2) -- (4,2) -- (4,6) -- (5,6) -- (5,1)
        (5,5) -- (6,5) -- (6,6)
        (5,4) -- (7,4) -- (7,6)
        (5,3) -- (8,3) -- (8,6)
        (5,2) -- (9,2) -- (9,6);

    \draw[blue, ultra thick]
        (0.5, 5.5) -- (1.5, 5.5) -- (1.5,1.5) -- (4.5,1.5) -- (4.5, 5.5)
        (2.5, 6.5) -- (1.5, 6.5) -- (1.5, 9.5) -- (9.5, 9.5) -- (9.5, 6.5) -- (5.5 , 6.5)
        (5.5, 1.5) -- (9.5,1.5) -- (9.5, 5.5)
        (2.5, 3.5) -- (2.5, 2.5) -- (3.5, 2.5) -- (3.5, 3.5)
        (7.5, 8.5) -- (8.5, 8.5) -- (8.5, 7.5) -- (7.5, 7.5)
        (3.5, 8.5) -- (2.5, 8.5) -- (2.5,7.5)
        (5.5, 2.5) -- (8.5, 2.5) -- (8.5, 5.5)
        (5.5, 3.5) -- (7.5, 3.5) -- (7.5, 5.5)
        (5.5, 4.5) -- (6.5, 4.5) -- (6.5, 5.5);

    \draw[red, ultra thick]
        (2.5, 5.5) -- (2.5, 3.5)
        (3.5, 3.5) -- (3.5, 5.5)
        (5.5, 8.5) -- (7.5, 8.5)
        (7.5, 7.5) -- (5.5, 7.5);

\end{tikzpicture}
\caption{Step 3}
\end{subfigure}

\vspace{1em}

\begin{subfigure}[t]{0.45\textwidth}
\centering
\begin{tikzpicture}[scale=0.4]
    \draw[step=1cm,gray,very thin] (0,0) grid (11,11);
    \draw[violet, ultra thick] (1,1) rectangle (10,10);

    \draw[violet, ultra thick]
        (1,6) -- (3,6) -- (3,7) -- (2,7) -- (2,9) -- (9,9) -- (9,7) -- (5,7) -- (5,6) -- (10,6)
        (2,6) -- (2,2) -- (4,2) -- (4,6) -- (5,6) -- (5,1)
        (5,5) -- (6,5) -- (6,6)
        (5,4) -- (7,4) -- (7,6)
        (5,3) -- (8,3) -- (8,6)
        (5,2) -- (9,2) -- (9,6);

    \draw[blue, ultra thick]
        (0.5, 5.5) -- (1.5, 5.5) -- (1.5,1.5) -- (4.5,1.5) -- (4.5, 5.5)
        (2.5, 6.5) -- (1.5, 6.5) -- (1.5, 9.5) -- (9.5, 9.5) -- (9.5, 6.5) -- (5.5 , 6.5)
        (5.5, 1.5) -- (9.5,1.5) -- (9.5, 5.5)
        (2.5, 5.5) -- (2.5, 2.5) -- (3.5, 2.5) -- (3.5, 5.5)
        (5.5, 8.5) -- (8.5, 8.5) -- (8.5, 7.5) -- (5.5, 7.5)
        (3.5, 8.5) -- (2.5, 8.5) -- (2.5,7.5)
        (5.5, 2.5) -- (8.5, 2.5) -- (8.5, 5.5)
        (5.5, 3.5) -- (7.5, 3.5) -- (7.5, 5.5)
        (5.5, 4.5) -- (6.5, 4.5) -- (6.5, 5.5);

    \draw[red, ultra thick]
        (4.5, 8.5) -- (5.5, 8.5)
        (2.5, 6.5) -- (2.5, 5.5);

\end{tikzpicture}

\caption{Step 4}

\end{subfigure}
\hfill
\begin{subfigure}[t]{0.45\textwidth}
\centering
\begin{tikzpicture}[scale=0.4]

    \draw[step=1cm,gray,very thin] (0,0) grid (11,11);
    \draw[violet, ultra thick] (1,1) rectangle (10,10);

    \draw[violet, ultra thick]
        (1,6) -- (3,6) -- (3,7) -- (2,7) -- (2,9) -- (9,9) -- (9,7) -- (5,7) -- (5,6) -- (10,6)
        (2,6) -- (2,2) -- (4,2) -- (4,6) -- (5,6) -- (5,1)
        (5,5) -- (6,5) -- (6,6)
        (5,4) -- (7,4) -- (7,6)
        (5,3) -- (8,3) -- (8,6)
        (5,2) -- (9,2) -- (9,6);

    \draw[blue, ultra thick]
        (0.5, 5.5) -- (1.5, 5.5) -- (1.5,1.5) -- (4.5,1.5) -- (4.5, 5.5)
        (2.5, 6.5) -- (1.5, 6.5) -- (1.5, 9.5) -- (9.5, 9.5) -- (9.5, 6.5) -- (5.5 , 6.5)
        (5.5, 1.5) -- (9.5,1.5) -- (9.5, 5.5)
        (2.5, 6.5) -- (2.5, 2.5) -- (3.5, 2.5) -- (3.5, 5.5)
        (4.5, 8.5) -- (8.5, 8.5) -- (8.5, 7.5) -- (5.5, 7.5)
        (3.5, 8.5) -- (2.5, 8.5) -- (2.5,7.5)
        (5.5, 2.5) -- (8.5, 2.5) -- (8.5, 5.5)
        (5.5, 3.5) -- (7.5, 3.5) -- (7.5, 5.5)
        (5.5, 4.5) -- (6.5, 4.5) -- (6.5, 5.5);

    \draw[red, ultra thick]
        (2.5,7.5) -- (3.5,7.5)
        (3.5, 8.5) -- (4.5, 8.5);

\end{tikzpicture}
\caption{Step 5}
\end{subfigure}

\vspace{1em}

\begin{subfigure}[t]{0.45\textwidth}
\centering
\begin{tikzpicture}[scale=0.4]
    \draw[step=1cm,gray,very thin] (0,0) grid (11,11);
    \draw[violet, ultra thick] (1,1) rectangle (10,10);

    \draw[violet, ultra thick]
        (1,6) -- (3,6) -- (3,7) -- (2,7) -- (2,9) -- (9,9) -- (9,7) -- (5,7) -- (5,6) -- (10,6)
        (2,6) -- (2,2) -- (4,2) -- (4,6) -- (5,6) -- (5,1)
        (5,5) -- (6,5) -- (6,6)
        (5,4) -- (7,4) -- (7,6)
        (5,3) -- (8,3) -- (8,6)
        (5,2) -- (9,2) -- (9,6);

    \draw[blue, ultra thick]
        (0.5, 5.5) -- (1.5, 5.5) -- (1.5,1.5) -- (4.5,1.5) -- (4.5, 5.5)
        (2.5, 6.5) -- (1.5, 6.5) -- (1.5, 9.5) -- (9.5, 9.5) -- (9.5, 6.5) -- (5.5 , 6.5)
        (5.5, 1.5) -- (9.5,1.5) -- (9.5, 5.5)
        (2.5, 6.5) -- (2.5, 2.5) -- (3.5, 2.5) -- (3.5, 5.5)
        (4.5, 8.5) -- (8.5, 8.5) -- (8.5, 7.5) -- (5.5, 7.5)
        (3.5, 8.5) -- (2.5, 8.5) -- (2.5,7.5) -- (3.5,7.5)
        (5.5, 2.5) -- (8.5, 2.5) -- (8.5, 5.5)
        (5.5, 3.5) -- (7.5, 3.5) -- (7.5, 5.5)
        (5.5, 4.5) -- (6.5, 4.5) -- (6.5, 5.5)
        (3.5, 8.5) -- (4.5, 8.5);

    \draw[red, ultra thick]
        (5.5 , 6.5) -- (5.5, 5.5);

\end{tikzpicture}

\caption{Step 6}

\end{subfigure}
\hfill
\begin{subfigure}[t]{0.45\textwidth}
\centering
\begin{tikzpicture}[scale=0.4]

    \draw[step=1cm,gray,very thin] (0,0) grid (11,11);
    \draw[violet, ultra thick] (1,1) rectangle (10,10);

    \draw[violet, ultra thick]
        (1,6) -- (3,6) -- (3,7) -- (2,7) -- (2,9) -- (9,9) -- (9,7) -- (5,7) -- (5,6) -- (10,6)
        (2,6) -- (2,2) -- (4,2) -- (4,6) -- (5,6) -- (5,1)
        (5,5) -- (6,5) -- (6,6)
        (5,4) -- (7,4) -- (7,6)
        (5,3) -- (8,3) -- (8,6)
        (5,2) -- (9,2) -- (9,6);

    \draw[blue, ultra thick]
        (0.5, 5.5) -- (1.5, 5.5) -- (1.5,1.5) -- (4.5,1.5) -- (4.5, 5.5)
        (2.5, 6.5) -- (1.5, 6.5) -- (1.5, 9.5) -- (9.5, 9.5) -- (9.5, 6.5) -- (5.5 , 6.5) -- (5.5, 5.5)
        (5.5, 1.5) -- (9.5,1.5) -- (9.5, 5.5)
        (2.5, 6.5) -- (2.5, 2.5) -- (3.5, 2.5) -- (3.5, 5.5)
        (4.5, 8.5) -- (8.5, 8.5) -- (8.5, 7.5) -- (5.5, 7.5)
        (3.5, 8.5) -- (2.5, 8.5) -- (2.5,7.5) -- (3.5,7.5)
        (3.5, 8.5) -- (4.5, 8.5)
        (5.5, 2.5) -- (8.5, 2.5) -- (8.5, 5.5)
        (5.5, 3.5) -- (7.5, 3.5) -- (7.5, 5.5)
        (5.5, 4.5) -- (6.5, 4.5) -- (6.5, 5.5);

    \draw[red, ultra thick]
        (5.5, 7.5) -- (4.5, 7.5) -- (4.5, 5.5)
        (3.5, 7.5) -- (3.5, 5.5);

\end{tikzpicture}
\caption{Step 7}
\end{subfigure}
\caption{9x9 forced-edge undirected T-metacell for Entry-Exit step-by-step traversal}
\label{6}
\end{figure}

\begin{figure}[H]
\centering

\begin{subfigure}[t]{0.45\textwidth}
\centering
\begin{tikzpicture}[scale=0.4]
    \draw[step=1cm,gray,very thin] (0,0) grid (11,11);
    \draw[violet, ultra thick] (1,1) rectangle (10,10);

    \draw[violet, ultra thick]
        (1,6) -- (3,6) -- (3,7) -- (2,7) -- (2,9) -- (9,9) -- (9,7) -- (5,7) -- (5,6) -- (10,6)
        (2,6) -- (2,2) -- (4,2) -- (4,6) -- (5,6) -- (5,1)
        (5,5) -- (6,5) -- (6,6)
        (5,4) -- (7,4) -- (7,6)
        (5,3) -- (8,3) -- (8,6)
        (5,2) -- (9,2) -- (9,6);

    \draw[blue, ultra thick]
        (0.5, 5.5) -- (1.5, 5.5) -- (1.5,1.5) -- (4.5,1.5) -- (4.5, 5.5)
        (2.5, 6.5) -- (1.5, 6.5) -- (1.5, 9.5) -- (9.5, 9.5) -- (9.5, 6.5) -- (5.5 , 6.5) -- (5.5, 5.5)
        (5.5, 1.5) -- (9.5,1.5) -- (9.5, 5.5)
        (2.5, 6.5) -- (2.5, 2.5) -- (3.5, 2.5) -- (3.5, 5.5) -- (3.5, 7.5)
        (4.5, 8.5) -- (8.5, 8.5) -- (8.5, 7.5) -- (5.5, 7.5) -- (4.5, 7.5) -- (4.5, 5.5)
        (3.5, 8.5) -- (2.5, 8.5) -- (2.5,7.5) -- (3.5,7.5)
        (3.5, 8.5) -- (4.5, 8.5)
        (5.5, 2.5) -- (8.5, 2.5) -- (8.5, 5.5)
        (5.5, 3.5) -- (7.5, 3.5) -- (7.5, 5.5)
        (5.5, 4.5) -- (6.5, 4.5) -- (6.5, 5.5)
        (5.5, 5.5) -- (5.5, 4.5)
        (6.5, 5.5) -- (7.5, 5.5)
        (5.5, 3.5) -- (5.5, 2.5)
        (8.5, 5.5) -- (9.5, 5.5)
        (5.5, 1.5) -- (5.5, 0.5);

\end{tikzpicture}

\caption{Filled T-metacell variant 1}

\end{subfigure}
\hfill
\begin{subfigure}[t]{0.45\textwidth}
\centering
\begin{tikzpicture}[scale=0.4]

    \draw[step=1cm,gray,very thin] (0,0) grid (11,11);
    \draw[violet, ultra thick] (1,1) rectangle (10,10);

    \draw[violet, ultra thick]
        (1,6) -- (3,6) -- (3,7) -- (2,7) -- (2,9) -- (9,9) -- (9,7) -- (5,7) -- (5,6) -- (10,6)
        (2,6) -- (2,2) -- (4,2) -- (4,6) -- (5,6) -- (5,1)
        (5,5) -- (6,5) -- (6,6)
        (5,4) -- (7,4) -- (7,6)
        (5,3) -- (8,3) -- (8,6)
        (5,2) -- (9,2) -- (9,6);

    \draw[blue, ultra thick]
        (0.5, 5.5) -- (1.5, 5.5) -- (1.5,1.5) -- (4.5,1.5) -- (4.5, 5.5)
        (2.5, 6.5) -- (1.5, 6.5) -- (1.5, 9.5) -- (9.5, 9.5) -- (9.5, 6.5) -- (5.5 , 6.5) -- (5.5, 5.5)
        (5.5, 1.5) -- (9.5,1.5) -- (9.5, 5.5)
        (2.5, 6.5) -- (2.5, 2.5) -- (3.5, 2.5) -- (3.5, 5.5) -- (3.5, 7.5)
        (4.5, 8.5) -- (8.5, 8.5) -- (8.5, 7.5) -- (5.5, 7.5) -- (4.5, 7.5) -- (4.5, 5.5)
        (3.5, 8.5) -- (2.5, 8.5) -- (2.5,7.5) -- (3.5,7.5)
        (3.5, 8.5) -- (4.5, 8.5)
        (5.5, 2.5) -- (8.5, 2.5) -- (8.5, 5.5)
        (5.5, 3.5) -- (7.5, 3.5) -- (7.5, 5.5)
        (5.5, 4.5) -- (6.5, 4.5) -- (6.5, 5.5)
        (5.5, 5.5) -- (6.5, 5.5)
        (5.5, 4.5) -- (5.5, 3.5)
        (7.5, 5.5) -- (8.5, 5.5)
        (5.5, 2.5) -- (5.5, 1.5)
        (9.5, 5.5) -- (10.5, 5.5);

\end{tikzpicture}
\caption{Filled T-metacell variant 2}
\end{subfigure}
   \label{7}
\end{figure}

Due to the uniqueness of the traversal, this proves that Entry-Exit is ASP-complete.

\subsection{Zahlenschlange}
In Zahlenschlange, a rectangular grid is filled with numbers, not necessarily unique, and the goal is to draw a line from the top left corner to the bottom right corner such that each unique number appears exactly once on the line. 

\begin{figure}[H]
    \centering
    \begin{subfigure}[t]{0.45\textwidth}
    \centering
    \begin{tikzpicture}[scale=0.7]
    \def\numbers{
        {5,4,3,9,1,2},
        {2,3,8,4,12,4},
        {12,9,6,8,7,1},
        {7,15,1,8,4,3},
        {3,8,1,2,6,1},
        {6,9,12,3,12,7},
    }

    \foreach \row [count=\y from 0] in \numbers {
        \foreach \num [count=\x from 0] in \row {
            \draw (\x,-\y) rectangle ++(1,-1);
            \node at (\x+0.5,-\y-0.5) {\num};
        }
    }
        
\end{tikzpicture}
    \caption{Unfilled Example}
    \end{subfigure}
    \hfill
    \begin{subfigure}[t]{0.45\textwidth}
    \centering
        \begin{tikzpicture}[scale=0.7]
    \def\numbers{
        {5,4,3,9,1,2},
        {2,3,8,4,12,4},
        {12,9,6,8,7,1},
        {7,15,1,8,4,3},
        {3,8,1,2,6,1},
        {6,9,12,3,12,7},
    }

    \foreach \row [count=\y from 0] in \numbers {
        \foreach \num [count=\x from 0] in \row {
            \draw (\x,-\y) rectangle ++(1,-1);
            \node at (\x+0.5,-\y-0.5) {\num};
        }
    }

    \draw[blue, ultra thick]
        (0.5, -0.5) -- (0.5, -1.5) -- (1.5, -1.5) -- (1.5, -3.5) -- (4.5, -3.5) -- (4.5, -5.5) -- (5.5, -5.5);
        
\end{tikzpicture}
    \caption{Filled Example}
    \end{subfigure}
		\label{8}
\end{figure}

More examples can be found at \cite{Zahlenschlange2025}.

This puzzle requires additional work beyond the T-metacell design. First, for each T-metacell, designate a unique non-negative integer n and construct the T-metacell in the way shown in \cref{9}.

\begin{figure}[H]
    \centering
    \begin{tikzpicture}[scale=1.3]
    \def\numbers{
        {0,0,0},
        {$3n+ 1$,$3n+ 2$,$3n+ 3$},
        {0,$3n+ 3$, 0},
    }

    \foreach \row [count=\y from 0] in \numbers {
        \foreach \num [count=\x from 0] in \row {
            \draw (\x,-\y) rectangle ++(1,-1);
            \node at (\x+0.5,-\y-0.5) {\num};
        }
    }
        
\end{tikzpicture}
    \caption{$n^{th}$ Asymmetrical Forced-edge Undirected T-metacell}
		\label{9}
\end{figure}

Assuming the number 0 was already used then the line may not contain it again, forcing the line to use either the right or the bottom exit. By design, $3n + 1$ and $3n + 2$ will only appear once in the entire grid and must therefore be used, effectively making the left exit the forced one. 

Next we modify the overall structure. Since the goal of the T-metacell framework is to build a Hamiltonian cycle, the top left corner can be chosen as the starting and ending point, allowing the use of the special T-metacell in \cref{10}. Assume that the top left cell uses $n = 0$ and that $n_i$ for $i \in \mathbb{N}$ is an arbitrary number not used by any T-metacells.

\begin{figure}[H]
    \centering
    \begin{tikzpicture}[scale=0.8]
    \def\numbers{
        {0,0,0},
        {1,2,3},
        {$n_2$, $n_1$, 0},
    }

    \foreach \row [count=\y from 0] in \numbers {
        \foreach \num [count=\x from 0] in \row {
            \draw (\x,-\y) rectangle ++(1,-1);
            \node at (\x+0.5,-\y-0.5) {\num};
        }
    }
        
\end{tikzpicture}
    \caption{Top-left corner cell}
		\label{10}
\end{figure}

First, add another row and column of cells filled with 0s except for the top left corner. Next, add another column to the left. Note that the top left corner is a 0, so the line may not use it again in any T-metacells. Connect the new starting point to the top left T-metacell. Finally, add a row at the bottom of the graph, along with a trail of arbitrary unused numbers from the top left T-metacell to the end. An example is shown in \cref{11}. The only entrance and exit of the T-metacells are at the top left one, so the path through the cells must therefore form a Hamiltonian cycle. 

\begin{figure}[H]
    \centering
    \begin{tikzpicture}[scale=0.8]
    \def\numbers{
        {0, $n_3$,0,0,0,0,0,0},
        {0, $n_4$,1,2,3,4,5,6},
        {$n_6$, $n_5$, $n_2$, $n_1$, 0, 0 ,6, 0},
        {$n_7$, 0,0,12,0,0,7,0},
        {$n_8$,0, 12, 11, 10, 9, 8, 0},
        {$n_9$,0, 0, 0, 0, 0, 9, 0},
        {$n_{10}$,0, 0, 0, 0, 0, 0, 0},
        {$n_{11}$,$n_{12}$,$n_{13}$,$n_{14}$,$n_{15}$,$n_{16}$,$n_{17}$,$n_{18}$},
    }

    \foreach \row [count=\y from 0] in \numbers {
        \foreach \num [count=\x from 0] in \row {
            \draw (\x,-\y) rectangle ++(1,-1);
            \node at (\x+0.5,-\y-0.5) {\num};
        }
    }
    \draw[black, ultra thick] (0,0) rectangle (8,-8);
    
    \draw[blue, ultra thick, dashed]
        (2,0) -- (2,-6) -- (8,-6);
        
\end{tikzpicture}
    \caption{Modified example Zahlenschlange instance}
		\label{11}
\end{figure}

The path from the top left corner to the starting cell is unique, the path to traverse each pair of exit of a cell is unique, and the path from the starting cell to the bottom right corner is also unique. Therefore, this proves the ASP-completeness of Zahlenschalnge.

A possible variant of this puzzle could require the line to go back to the top left cell. It is easy to see that a similar reduction will work, with a small change of having the path after exiting the T-metacells lead back to the top left instead.

\clearpage

\subsection{Yagit}
In Yagit, we have a grid in which some cells are filled with either a sheep or a wolf. Each grid intersection point may have a black dot. The goal is to partition the grid into regions using any number of lines such that each region must contain at least one sheep or at least one wolf, but not both. Each partition lines must start and end at unique points on the border of the grid and may only turn 90 degrees at a black dot, but they are not required to do so. The partition lines can intersect, but only where there are no black dots. Not all of the black dots must be used. For simplicity, sheep are represented with black dots, and wolves are represented with red dots.

Although the rules are a bit complicated, there are possible parallels to this problem in real life such as electoral zoning.

\begin{figure}[H]
    \begin{subfigure}[t]{0.45\textwidth}
    \centering
        \begin{tikzpicture}[scale=0.8]
    \draw[step=1cm,gray,very thin] (0,0) grid (6,6);
    \draw[line width=1pt] (0,0) rectangle (6,6);

    \def\blackcells{0/5, 5/5, 3/4, 3/2, 2/2, 0/1, 4/0}
    \def\redcells{0/2, 0/3, 2/0, 2/5, 3/3, 4/3, 5/2, 5/3, 5/4}
    \def\blacklines{0/2, 1/3, 2/1}

    \foreach \x/\y in \blackcells {
        \fill[black] (\x+0.5,\y+0.5) circle (0.15cm);
    }

    \foreach \x/\y in \redcells {
        \fill[red] (\x+0.5,\y+0.5) circle (0.15cm);
    }

    \foreach \x/\y in \blacklines {
        \fill[black] (\x+1,\y+1) circle (0.15cm);
    }

\end{tikzpicture}
    \caption{Unfilled Example}
    \end{subfigure}
    \hfill
    \begin{subfigure}[t]{0.45\textwidth}
    \centering
        \begin{tikzpicture}[scale=0.8]
        \draw[step=1cm,gray,very thin] (0,0) grid (6,6);
        \draw[line width=1pt] (0,0) rectangle (6,6);
    
        \def\blackcells{0/5, 5/5, 3/4, 3/2, 2/2, 0/1, 4/0}
        \def\redcells{0/2, 0/3, 2/0, 2/5, 3/3, 4/3, 5/2, 5/3, 5/4}
        \def\blacklines{0/2, 1/3, 2/1}
    
        \foreach \x/\y in \blackcells {
            \fill[black] (\x+0.5,\y+0.5) circle (0.15cm);
        }
    
        \foreach \x/\y in \redcells {
            \fill[red] (\x+0.5,\y+0.5) circle (0.15cm);
        }
    
        \foreach \x/\y in \blacklines {
            \fill[black] (\x+1,\y+1) circle (0.15cm);
        }
    
        \draw[blue, ultra thick]
            (0,5) -- (6,5)
            (2,6) -- (2,4) -- (6,4)
            (0,2) -- (3,2) -- (3,0)
            (1,0) -- (1,3) -- (6,3)
            (5,6) -- (5,0);

        \end{tikzpicture}
        \caption{Filled Example}
        \end{subfigure}
				\label{12}
\end{figure}

More examples can be found at \cite{Yagit2025}.

\begin{figure}[H]
    \centering
    \begin{tikzpicture}[scale=0.8]
    \draw[step=1cm,gray,very thin] (0,0) grid (3,3);
    \draw[line width=1pt] (0,0) rectangle (3,3);

    \def\blackcells{0/1, 1/1}
    \def\redcells{0/2, 1/2, 2/2, 0/0, 2/0}
    \def\blacklines{0/0, 0/1, 1/1}

    \foreach \x/\y in \blackcells {
        \fill[black] (\x+0.5,\y+0.5) circle (0.15cm);
    }

    \foreach \x/\y in \redcells {
        \fill[red] (\x+0.5,\y+0.5) circle (0.15cm);
    }

    \foreach \x/\y in \blacklines {
        \fill[black] (\x+1,\y+1) circle (0.15cm);
    }

\end{tikzpicture}
    \caption{Asymmetrical Forced-edge Undirected T-metacell for Yagit}
		\label{13}
\end{figure}

A T-metacell for Yagit can be constructed as shown in \cref{13}, however, the proof for Yagit requires some more work. The main idea is that if we restrict the starting point of the borders to just one, then the T-metacell becomes more powerful. To reduce from Hamiltonian cycle in a rectangular grid of asymmetrical forced-edge undirected T-metacell, first, replace each cell with the one shown in the diagram in a similar way to the demonstration in \cref{14}.

\begin{figure}[H]
\centering

\begin{subfigure}[t]{0.45\textwidth}
\centering
\begin{tikzpicture}[scale=1]
    \def\w{1}
    \def\h{1}

    \draw[step=1cm,gray,thin] (0,0) grid (4,4);

    \newcommand{\drawJunction}[3]{
        \begin{scope}[shift={(#1+0.5,#2+0.5)}, rotate=#3]
            \draw[thick] (0,0) -- (0,0.5);
            \draw[line width=2pt] (0,0) -- (0.5,0);
            \draw[thick] (0,0) -- (-0.5,0);
        \end{scope}
    }

    \newcommand{\drawInvertedJunction}[3]{
        \begin{scope}[shift={(#1+0.5,#2+0.5)}, rotate=#3]
            \draw[thick] (0,0) -- (0,0.5);
            \draw[thick] (0,0) -- (0.5,0);
            \draw[line width=2pt] (0,0) -- (-0.5,0);
        \end{scope}
    }

    \drawJunction{0}{0}{0}
    \drawJunction{1}{0}{180}
    \drawJunction{2}{0}{0}
    \drawJunction{3}{0}{90}

    \drawInvertedJunction{0}{1}{180}
    \drawInvertedJunction{1}{1}{0}
    \drawInvertedJunction{2}{1}{270}
    \drawJunction{3}{1}{180}

    \drawInvertedJunction{0}{2}{270}
    \drawJunction{1}{2}{180}
    \drawJunction{2}{2}{270}
    \drawJunction{3}{2}{90}

    \drawInvertedJunction{0}{3}{180}
    \drawInvertedJunction{1}{3}{180}
    \drawJunction{2}{3}{0}
    \drawJunction{3}{3}{180}

\end{tikzpicture}

\caption{Original forced-edge T-metacell grid}
\end{subfigure}
\hfill
\begin{subfigure}[t]{0.45\textwidth}
\centering
   \begin{tikzpicture}[scale=0.5]
    \draw[step=1cm,gray,very thin] (0,0) grid (12,12);
    \draw[line width=1pt] (0,0) rectangle (12,12);

    \def\blackcells{1/1, 1/4, 1/7, 1/8, 1/10, 2/1, 2/4, 2/10, 3/1, 3/4, 3/7, 4/1, 4/4, 
4/7, 4/10, 5/10, 6/1, 7/1, 7/4, 7/5, 7/6, 7/7, 7/10, 8/10, 9/4, 9/10, 10/1, 10/2, 10/4, 10/7, 10/8, 10/10}
    \def\redcells{0/0, 0/2, 0/3, 0/5, 0/6, 0/7, 0/8, 0/9, 0/11, 1/0, 1/11, 2/0, 2/11, 3/0, 3/11, 4/11, 5/0, 5/11, 6/0, 6/11, 7/0, 8/0, 8/11, 9/0, 9/11, 10/11, 11/0, 11/1, 11/2, 11/3, 11/5, 11/6, 11/7, 11/8, 11/9, 11/11, 1/5, 2/2, 2/3, 2/5, 2/6, 2/8, 2/9, 3/2, 3/3, 3/5, 3/6, 3/8, 3/9, 4/2, 4/3, 4/8, 5/2, 5/3, 5/5, 5/6, 5/8, 5/9, 6/2, 6/3, 6/5, 6/6, 6/7, 
6/8, 6/9, 7/9, 8/2, 8/3, 8/5, 8/6, 8/8, 8/9, 9/2, 9/3, 9/5, 9/6, 9/8, 9/9, 10/5, 6/4}
    \def\blacklines{0/0, 0/4, 0/6, 0/7, 0/10, 1/0, 1/1, 1/3, 1/4, 1/7, 1/9, 1/10, 3/0, 
3/1, 3/3, 3/4, 3/6, 3/7, 3/10, 4/1, 4/3, 4/7, 4/9, 4/10, 6/0, 6/1, 6/3, 6/4, 6/6, 6/7, 6/9, 7/0, 7/4, 7/6, 7/9, 7/10, 9/1, 9/3, 9/4, 9/7, 9/9, 9/10, 10/0, 10/1, 10/4, 10/6, 10/7, 10/10}

    \foreach \x/\y in \blackcells {
        \fill[black] (\x+0.5,\y+0.5) circle (0.15cm);
    }

    \foreach \x/\y in \redcells {
        \fill[red] (\x+0.5,\y+0.5) circle (0.15cm);
    }

    \foreach \x/\y in \blacklines {
        \fill[black] (\x+1,\y+1) circle (0.15cm);
    }

    \draw[black, ultra thick, dashed]
        (3,0) -- (3,12)
        (6,0) -- (6,12)
        (9,0) -- (9,12)
        (0,3) -- (12,3)
        (0,6) -- (12,6)
        (0,9) -- (12,9)
        ;

\end{tikzpicture}
\caption{Corresponding Yagit instance}
\end{subfigure}
\caption{Example reduction from T-metacell grid}
\label{14}
\end{figure}

Then, for each T-metacell on the border, fill in impossible sides that faces the border with wolves. Since the ultimate goal is to construct a Hamiltonian cycle, any point can be chosen to be the starting and ending point. We can replace the top left cell with the special cell in \cref{15}.  An extra sheep is also added to the cell on the right of the top left cell, assuming that they are connected because the answer would be a trivial no otherwise.

\begin{figure}[H]
\centering

\begin{subfigure}[t]{0.45\textwidth}
\centering
\begin{tikzpicture}[scale=1]
    \draw[step=1cm,gray,very thin] (0,0) grid (3,3);
    \draw[line width=1pt] (0,0) rectangle (3,3);

    \def\blackcells{1/1}
    \def\redcells{0/2, 0/1, 0/0, 2/2, 2/1, 2/0}
    \def\blacklines{2/0, 2/1}

    \foreach \x/\y in \blackcells {
        \fill[black] (\x+0.5,\y+0.5) circle (0.15cm);
    }

    \foreach \x/\y in \redcells {
        \fill[red] (\x+0.5,\y+0.5) circle (0.15cm);
    }

    \foreach \x/\y in \blacklines {
        \fill[black] (\x+1,\y+1) circle (0.15cm);
    }

\end{tikzpicture}
\caption{Top-left corner cell}
\end{subfigure}
\hfill
\begin{subfigure}[t]{0.45\textwidth}
\centering
\begin{tikzpicture}[scale=0.5]
    \draw[step=1cm,gray,very thin] (0,0) grid (12,12);
    \draw[line width=1pt] (0,0) rectangle (12,12);

    \def\blackcells{1/1, 1/4, 1/7, 1/8, 1/10, 2/1, 2/4, 3/1, 3/4, 3/7, 4/1, 4/4, 
4/7, 4/10, 5/10, 6/1, 7/1, 7/4, 7/5, 7/6, 7/7, 7/10, 8/10, 9/4, 9/10, 10/1, 10/2, 10/4, 10/7, 10/8, 10/10, 3/10}
    \def\redcells{0/0, 0/1, 0/2, 0/3, 0/4, 0/5, 0/6, 0/7, 0/8, 0/9, 0/10, 0/11, 1/0, 2/0, 2/11, 3/0, 3/11, 4/0, 4/11, 5/0, 5/11, 6/0, 6/11, 7/0, 7/11, 8/0, 8/11, 9/0, 9/11, 10/0, 10/11, 11/0, 11/1, 11/2, 11/3, 11/4, 11/5, 11/6, 11/7, 11/8, 11/9, 11/10, 11/11, 1/5, 2/2, 2/3, 2/5, 2/6, 2/8, 2/9, 3/2, 3/3, 3/5, 3/6, 3/8, 3/9, 4/2, 4/3, 4/8, 5/2, 5/3, 5/5, 5/6, 5/8, 5/9, 6/2, 6/3, 6/5, 6/6, 6/7, 
6/8, 6/9, 7/9, 8/2, 8/3, 8/5, 8/6, 8/8, 8/9, 9/2, 9/3, 9/5, 9/6, 9/8, 9/9, 10/5, 6/4, 2/10}
    \def\blacklines{0/0, 0/4, 0/6, 0/7, 1/0, 1/1, 1/3, 1/4, 1/7, 3/0, 
3/1, 3/3, 3/4, 3/6, 3/7, 3/10, 4/1, 4/3, 4/7, 4/9, 4/10, 6/0, 6/1, 6/3, 6/4, 6/6, 6/7, 6/9, 7/0, 7/4, 7/6, 7/9, 7/10, 9/1, 9/3, 9/4, 9/7, 9/9, 9/10, 10/0, 10/1, 10/4, 10/6, 10/7, 10/10, 2/10, 2/9}

    \foreach \x/\y in \blackcells {
        \fill[black] (\x+0.5,\y+0.5) circle (0.15cm);
    }

    \foreach \x/\y in \redcells {
        \fill[red] (\x+0.5,\y+0.5) circle (0.15cm);
    }

    \foreach \x/\y in \blacklines {
        \fill[black] (\x+1,\y+1) circle (0.15cm);
    }

    \draw[black, ultra thick, dashed]
        (3,0) -- (3,12)
        (6,0) -- (6,12)
        (9,0) -- (9,12)
        (0,3) -- (12,3)
        (0,6) -- (12,6)
        (0,9) -- (12,9)
        ;

\end{tikzpicture}
\caption{Modified Yagit instance}
\end{subfigure}
\caption{Example reduction from T-metacell grid (continued)}
\label{15}
\end{figure}

Next, the whole structure is wrapped with a layer of sheep except for the top left corner as shown in \cref{16}.

\begin{figure}[H]
\centering

\begin{subfigure}[t]{0.45\textwidth}
\centering
\begin{tikzpicture}[scale=0.44]
    \draw[step=1cm,gray,very thin] (0,0) grid (14,14);
    \draw[line width=1pt] (0,0) rectangle (14,14);

    \def\blackcells{0/0, 0/1, 0/2, 0/3, 0/4, 0/5, 0/6, 0/7, 0/8, 0/9, 0/10, 0/11, 0/12, 0/13, 1/0, 2/0, 3/0, 4/0, 4/13, 5/0, 5/13, 6/0, 6/13, 7/0, 7/13, 8/0, 8/13, 9/0, 9/13, 10/0, 10/13, 11/0, 11/13, 12/0, 12/13, 13/0, 13/1, 13/2, 13/3, 13/4, 13/5, 13/6, 13/7, 13/8, 13/9, 13/10, 13/11, 13/12, 13/13}
    \def\redcells{1/13, 3/13}
    \def\blacklines{0/0, 0/1, 0/2, 0/3, 0/4, 0/5, 0/6, 0/7, 0/8, 0/9, 0/10, 0/11, 0/12, 1/0, 2/0, 3/0, 3/12, 4/0, 4/12, 5/0, 5/12, 6/0, 6/12, 7/0, 7/12, 8/0, 8/12, 9/0, 9/12, 10/0, 10/12, 11/0, 11/12, 12/0, 12/1, 12/2, 12/3, 12/4, 12/5, 12/6, 12/7, 12/8, 12/9, 12/10, 12/11, 12/12}

    \foreach \x/\y in \blackcells {
        \fill[black] (\x+0.5,\y+0.5) circle (0.15cm);
    }

    \foreach \x/\y in \redcells {
        \fill[red] (\x+0.5,\y+0.5) circle (0.15cm);
    }

    \foreach \x/\y in \blacklines {
        \fill[black] (\x+1,\y+1) circle (0.15cm);
    }

\end{tikzpicture}
\caption{Additional cells around the original graph}
\end{subfigure}
\hfill
\begin{subfigure}[t]{0.45\textwidth}
\centering
\begin{tikzpicture}[scale=0.44]
    \draw[step=1cm,gray,very thin] (-1,-1) grid (13,13);
    \draw[line width=1pt] (-1,-1) rectangle (13,13);

    \def\blackcells{1/1, 1/4, 1/7, 1/8, 1/10, 2/1, 2/4, 3/1, 3/4, 3/7, 4/1, 4/4, 
4/7, 4/10, 5/10, 6/1, 7/1, 7/4, 7/5, 7/6, 7/7, 7/10, 8/10, 9/4, 9/10, 10/1, 10/2, 10/4, 10/7, 10/8, 10/10, 3/10, -1/-1, -1/0, -1/1, -1/2, -1/3, -1/4, -1/5, -1/6, -1/7, -1/8, -1/9, -1/10, -1/11, -1/12, 0/-1, 1/-1, 2/-1, 3/-1, 3/12, 4/-1, 4/12, 5/-1, 5/12, 6/-1, 6/12, 7/-1, 7/12, 8/-1, 8/12, 9/-1, 9/12, 10/-1, 10/12, 11/-1, 11/12, 12/-1, 12/0, 12/1, 12/2, 12/3, 12/4, 12/5, 12/6, 12/7, 12/8, 12/9, 12/10, 12/11, 12/12}
    \def\redcells{0/0, 0/1, 0/2, 0/3, 0/4, 0/5, 0/6, 0/7, 0/8, 0/9, 0/10, 0/11, 1/0, 2/0, 2/11, 3/0, 3/11, 4/0, 4/11, 5/0, 5/11, 6/0, 6/11, 7/0, 7/11, 8/0, 8/11, 9/0, 9/11, 10/0, 10/11, 11/0, 11/1, 11/2, 11/3, 11/4, 11/5, 11/6, 11/7, 11/8, 11/9, 11/10, 11/11, 1/5, 2/2, 2/3, 2/5, 2/6, 2/8, 2/9, 3/2, 3/3, 3/5, 3/6, 3/8, 3/9, 4/2, 4/3, 4/8, 5/2, 5/3, 5/5, 5/6, 5/8, 5/9, 6/2, 6/3, 6/5, 6/6, 6/7, 
6/8, 6/9, 7/9, 8/2, 8/3, 8/5, 8/6, 8/8, 8/9, 9/2, 9/3, 9/5, 9/6, 9/8, 9/9, 10/5, 6/4, 2/10, 0/12, 2/12}
    \def\blacklines{0/0, 0/4, 0/6, 0/7, 1/0, 1/1, 1/3, 1/4, 1/7, 3/0, 
3/1, 3/3, 3/4, 3/6, 3/7, 3/10, 4/1, 4/3, 4/7, 4/9, 4/10, 6/0, 6/1, 6/3, 6/4, 6/6, 6/7, 6/9, 7/0, 7/4, 7/6, 7/9, 7/10, 9/1, 9/3, 9/4, 9/7, 9/9, 9/10, 10/0, 10/1, 10/4, 10/6, 10/7, 10/10, 2/10, 2/9, -1/-1, -1/0, -1/1, -1/2, -1/3, -1/4, -1/5, -1/6, -1/7, -1/8, -1/9, -1/10, -1/11, 0/-1, 1/-1, 2/-1, 2/11, 3/-1, 3/11, 4/-1, 4/11, 5/-1, 5/11, 6/-1, 6/11, 7/-1, 7/11, 8/-1, 8/11, 9/-1, 9/11, 10/-1, 10/11, 11/-1, 11/0, 11/1, 11/2, 11/3, 11/4, 11/5, 11/6, 11/7, 11/8, 11/9, 11/10, 11/11}

    \foreach \x/\y in \blackcells {
        \fill[black] (\x+0.5,\y+0.5) circle (0.15cm);
    }

    \foreach \x/\y in \redcells {
        \fill[red] (\x+0.5,\y+0.5) circle (0.15cm);
    }

    \foreach \x/\y in \blacklines {
        \fill[black] (\x+1,\y+1) circle (0.15cm);
    }

    \draw[black, ultra thick, dashed]
        (3,0) -- (3,12)
        (6,0) -- (6,12)
        (9,0) -- (9,12)
        (0,3) -- (12,3)
        (0,6) -- (12,6)
        (0,9) -- (12,9)
        ;

\end{tikzpicture}
\caption{Modified Yagit instance}
\end{subfigure}

\vspace{1em}

\begin{subfigure}[t]{0.45\textwidth}
\centering
\begin{tikzpicture}[scale=0.44]
    \draw[step=1cm,gray,very thin] (-1,-1) grid (13,13);
    \draw[line width=1pt] (-1,-1) rectangle (13,13);

    \def\blackcells{1/1, 1/4, 1/7, 1/8, 1/10, 2/1, 2/4, 3/1, 3/4, 3/7, 4/1, 4/4, 
4/7, 4/10, 5/10, 6/1, 7/1, 7/4, 7/5, 7/6, 7/7, 7/10, 8/10, 9/4, 9/10, 10/1, 10/2, 10/4, 10/7, 10/8, 10/10, 3/10, -1/-1, -1/0, -1/1, -1/2, -1/3, -1/4, -1/5, -1/6, -1/7, -1/8, -1/9, -1/10, -1/11, -1/12, 0/-1, 1/-1, 2/-1, 3/-1, 3/12, 4/-1, 4/12, 5/-1, 5/12, 6/-1, 6/12, 7/-1, 7/12, 8/-1, 8/12, 9/-1, 9/12, 10/-1, 10/12, 11/-1, 11/12, 12/-1, 12/0, 12/1, 12/2, 12/3, 12/4, 12/5, 12/6, 12/7, 12/8, 12/9, 12/10, 12/11, 12/12}
    \def\redcells{0/0, 0/1, 0/2, 0/3, 0/4, 0/5, 0/6, 0/7, 0/8, 0/9, 0/10, 0/11, 1/0, 2/0, 2/11, 3/0, 3/11, 4/0, 4/11, 5/0, 5/11, 6/0, 6/11, 7/0, 7/11, 8/0, 8/11, 9/0, 9/11, 10/0, 10/11, 11/0, 11/1, 11/2, 11/3, 11/4, 11/5, 11/6, 11/7, 11/8, 11/9, 11/10, 11/11, 1/5, 2/2, 2/3, 2/5, 2/6, 2/8, 2/9, 3/2, 3/3, 3/5, 3/6, 3/8, 3/9, 4/2, 4/3, 4/8, 5/2, 5/3, 5/5, 5/6, 5/8, 5/9, 6/2, 6/3, 6/5, 6/6, 6/7, 
6/8, 6/9, 7/9, 8/2, 8/3, 8/5, 8/6, 8/8, 8/9, 9/2, 9/3, 9/5, 9/6, 9/8, 9/9, 10/5, 6/4, 2/10, 0/12, 2/12}
    \def\blacklines{0/0, 0/4, 0/6, 0/7, 1/0, 1/1, 1/3, 1/4, 1/7, 3/0, 
3/1, 3/3, 3/4, 3/6, 3/7, 3/10, 4/1, 4/3, 4/7, 4/9, 4/10, 6/0, 6/1, 6/3, 6/4, 6/6, 6/7, 6/9, 7/0, 7/4, 7/6, 7/9, 7/10, 9/1, 9/3, 9/4, 9/7, 9/9, 9/10, 10/0, 10/1, 10/4, 10/6, 10/7, 10/10, 2/10, 2/9, -1/-1, -1/0, -1/1, -1/2, -1/3, -1/4, -1/5, -1/6, -1/7, -1/8, -1/9, -1/10, -1/11, 0/-1, 1/-1, 2/-1, 2/11, 3/-1, 3/11, 4/-1, 4/11, 5/-1, 5/11, 6/-1, 6/11, 7/-1, 7/11, 8/-1, 8/11, 9/-1, 9/11, 10/-1, 10/11, 11/-1, 11/0, 11/1, 11/2, 11/3, 11/4, 11/5, 11/6, 11/7, 11/8, 11/9, 11/10, 11/11}

    \foreach \x/\y in \blackcells {
        \fill[black] (\x+0.5,\y+0.5) circle (0.15cm);
    }

    \foreach \x/\y in \redcells {
        \fill[red] (\x+0.5,\y+0.5) circle (0.15cm);
    }

    \foreach \x/\y in \blacklines {
        \fill[black] (\x+1,\y+1) circle (0.15cm);
    }

    \draw[black, ultra thick, dashed]
        (3,0) -- (3,12)
        (6,0) -- (6,12)
        (9,0) -- (9,12)
        (0,3) -- (12,3)
        (0,6) -- (12,6)
        (0,9) -- (12,9)
        ;

    \draw[blue, ultra thick]
        (0,13) -- (0,0) -- (12,0) -- (12,12) -- (3,12) -- (3,13)
        ;

\end{tikzpicture}
\caption{Modified instance with some forced edge shown}
\end{subfigure}

\caption{Example reduction from T-metacell grid (continued)}
\label{16}
\end{figure}

Clearly, there must be a border between a sheep and a wolf when they are adjacent. Using this and the fact that borders cannot intersect on black dots, the final structure is left with just two points on the border, forcing the partition to use exactly one line. For one border to partition all sheep and wolves in the original graph, the border must traverse every T-metacell and create a border between every adjacent wolves and sheep. 

Denote the two black dots on the right side of the top left cell as the turning point. If there is a path through all T-metacells that started from the entrance and ends here, then it represents a Hamiltonian cycle in the original graph. Notice that it is impossible to completely partition each cell by one line traversal. The following methods, shown in \cref{17}, are the only three ways for the border to traverse the T-metacell, denoted as type-a, type-b, and type-c. 

\begin{figure}[H]
\centering

\begin{subfigure}[t]{0.45\textwidth}
\centering
\begin{tikzpicture}[scale=1]
    \draw[step=1cm,gray,very thin] (0,0) grid (3,3);
    \draw[line width=1pt] (0,0) rectangle (3,3);

    \def\blackcells{0/1, 1/1}
    \def\redcells{0/2, 1/2, 2/2, 0/0, 2/0}
    \def\blacklines{0/0, 0/1, 1/1}

    \foreach \x/\y in \blackcells {
        \fill[black] (\x+0.5,\y+0.5) circle (0.15cm);
    }

    \foreach \x/\y in \redcells {
        \fill[red] (\x+0.5,\y+0.5) circle (0.15cm);
    }

    \foreach \x/\y in \blacklines {
        \fill[black] (\x+1,\y+1) circle (0.15cm);
    }

    \draw[blue, ultra thick]
        (0,2) -- (2,2)
        (0,1) -- (1,1);

\end{tikzpicture}
\caption{T-metacell with forced edges shown}
\end{subfigure}
\hfill
\begin{subfigure}[t]{0.45\textwidth}
\centering
\begin{tikzpicture}[scale=1]
    \draw[step=1cm,gray,very thin] (0,0) grid (3,3);
    \draw[line width=1pt] (0,0) rectangle (3,3);

    \def\blackcells{0/1, 1/1}
    \def\redcells{0/2, 1/2, 2/2, 0/0, 2/0}
    \def\blacklines{0/0, 0/1, 1/1}

    \foreach \x/\y in \blackcells {
        \fill[black] (\x+0.5,\y+0.5) circle (0.15cm);
    }

    \foreach \x/\y in \redcells {
        \fill[red] (\x+0.5,\y+0.5) circle (0.15cm);
    }

    \foreach \x/\y in \blacklines {
        \fill[black] (\x+1,\y+1) circle (0.15cm);
    }

    \draw[blue, ultra thick]
        (0,2) -- (3,2)
        (0,1) -- (3,1);

\end{tikzpicture}
\caption{T-metacell Traversal type-a}
\end{subfigure}

\vspace{1em}

\begin{subfigure}[t]{0.45\textwidth}
\centering
\begin{tikzpicture}[scale=1]
    \draw[step=1cm,gray,very thin] (0,0) grid (3,3);
    \draw[line width=1pt] (0,0) rectangle (3,3);

    \def\blackcells{0/1, 1/1}
    \def\redcells{0/2, 1/2, 2/2, 0/0, 2/0}
    \def\blacklines{0/0, 0/1, 1/1}

    \foreach \x/\y in \blackcells {
        \fill[black] (\x+0.5,\y+0.5) circle (0.15cm);
    }

    \foreach \x/\y in \redcells {
        \fill[red] (\x+0.5,\y+0.5) circle (0.15cm);
    }

    \foreach \x/\y in \blacklines {
        \fill[black] (\x+1,\y+1) circle (0.15cm);
    }

    \draw[blue, ultra thick]
        (0,2) -- (2,2) -- (2,0)
        (0,1) -- (1,1) -- (1,0);

\end{tikzpicture}
\caption{T-metacell Traversal type-b}
\end{subfigure}
\hfill
\begin{subfigure}[t]{0.45\textwidth}
\centering
\begin{tikzpicture}[scale=1]
    \draw[step=1cm,gray,very thin] (0,0) grid (3,3);
    \draw[line width=1pt] (0,0) rectangle (3,3);

    \def\blackcells{0/1, 1/1}
    \def\redcells{0/2, 1/2, 2/2, 0/0, 2/0}
    \def\blacklines{0/0, 0/1, 1/1}

    \foreach \x/\y in \blackcells {
        \fill[black] (\x+0.5,\y+0.5) circle (0.15cm);
    }

    \foreach \x/\y in \redcells {
        \fill[red] (\x+0.5,\y+0.5) circle (0.15cm);
    }

    \foreach \x/\y in \blacklines {
        \fill[black] (\x+1,\y+1) circle (0.15cm);
    }

    \draw[blue, ultra thick]
        (0,2) -- (2,2) -- (2,0)
        (0,1) -- (3,1);

\end{tikzpicture}
\caption{T-metacell Traversal type-c}
\end{subfigure}

\caption{All possible traversals of T-metacell}
\label{17}
\end{figure}

Next, we will show that given a solution to Yagit, we can construct a Hamiltonian cycle. If the whole solution of the puzzle consists of just type-a and type-b, then it is easy to see that the result will be one long region consisting of all the sheep starting at the entrance and ending at the turning point, representing the Hamiltonian cycle as shown in \cref{18}. A solution in the original T-metacell grid implies a solution in Yagit.

\begin{figure}[H]
    \centering
    \begin{subfigure}[t]{0.45\textwidth}
    \centering
    \begin{tikzpicture}[scale=0.44]
    \draw[step=1cm,gray,very thin] (-1,-1) grid (13,13);
    \draw[line width=1pt] (-1,-1) rectangle (13,13);

    \def\blackcells{1/1, 1/4, 1/7, 1/8, 1/10, 2/1, 2/4, 3/1, 3/4, 3/7, 4/1, 4/4, 
4/7, 4/10, 5/10, 6/1, 7/1, 7/4, 7/5, 7/6, 7/7, 7/10, 8/10, 9/4, 9/10, 10/1, 10/2, 10/4, 10/7, 10/8, 10/10, 3/10, -1/-1, -1/0, -1/1, -1/2, -1/3, -1/4, -1/5, -1/6, -1/7, -1/8, -1/9, -1/10, -1/11, -1/12, 0/-1, 1/-1, 2/-1, 3/-1, 3/12, 4/-1, 4/12, 5/-1, 5/12, 6/-1, 6/12, 7/-1, 7/12, 8/-1, 8/12, 9/-1, 9/12, 10/-1, 10/12, 11/-1, 11/12, 12/-1, 12/0, 12/1, 12/2, 12/3, 12/4, 12/5, 12/6, 12/7, 12/8, 12/9, 12/10, 12/11, 12/12}
    \def\redcells{0/0, 0/1, 0/2, 0/3, 0/4, 0/5, 0/6, 0/7, 0/8, 0/9, 0/10, 0/11, 1/0, 2/0, 2/11, 3/0, 3/11, 4/0, 4/11, 5/0, 5/11, 6/0, 6/11, 7/0, 7/11, 8/0, 8/11, 9/0, 9/11, 10/0, 10/11, 11/0, 11/1, 11/2, 11/3, 11/4, 11/5, 11/6, 11/7, 11/8, 11/9, 11/10, 11/11, 1/5, 2/2, 2/3, 2/5, 2/6, 2/8, 2/9, 3/2, 3/3, 3/5, 3/6, 3/8, 3/9, 4/2, 4/3, 4/8, 5/2, 5/3, 5/5, 5/6, 5/8, 5/9, 6/2, 6/3, 6/5, 6/6, 6/7, 
6/8, 6/9, 7/9, 8/2, 8/3, 8/5, 8/6, 8/8, 8/9, 9/2, 9/3, 9/5, 9/6, 9/8, 9/9, 10/5, 6/4, 2/10, 0/12, 2/12}
    \def\blacklines{0/0, 0/4, 0/6, 0/7, 1/0, 1/1, 1/3, 1/4, 1/7, 3/0, 
3/1, 3/3, 3/4, 3/6, 3/7, 3/10, 4/1, 4/3, 4/7, 4/9, 4/10, 6/0, 6/1, 6/3, 6/4, 6/6, 6/7, 6/9, 7/0, 7/4, 7/6, 7/9, 7/10, 9/1, 9/3, 9/4, 9/7, 9/9, 9/10, 10/0, 10/1, 10/4, 10/6, 10/7, 10/10, 2/10, 2/9, -1/-1, -1/0, -1/1, -1/2, -1/3, -1/4, -1/5, -1/6, -1/7, -1/8, -1/9, -1/10, -1/11, 0/-1, 1/-1, 2/-1, 2/11, 3/-1, 3/11, 4/-1, 4/11, 5/-1, 5/11, 6/-1, 6/11, 7/-1, 7/11, 8/-1, 8/11, 9/-1, 9/11, 10/-1, 10/11, 11/-1, 11/0, 11/1, 11/2, 11/3, 11/4, 11/5, 11/6, 11/7, 11/8, 11/9, 11/10, 11/11}

    \foreach \x/\y in \blackcells {
        \fill[black] (\x+0.5,\y+0.5) circle (0.15cm);
    }

    \foreach \x/\y in \redcells {
        \fill[red] (\x+0.5,\y+0.5) circle (0.15cm);
    }

    \foreach \x/\y in \blacklines {
        \fill[black] (\x+1,\y+1) circle (0.15cm);
    }

    \draw[black, ultra thick, dashed]
        (3,0) -- (3,12)
        (6,0) -- (6,12)
        (9,0) -- (9,12)
        (0,3) -- (12,3)
        (0,6) -- (12,6)
        (0,9) -- (12,9)
        ;

    \draw[blue, ultra thick]
        (0,13) -- (0,0) -- (12,0) -- (12,12) -- (3,12) -- (3,13)
        (1,13) -- (1,7) -- (4,7) -- (4,5) -- (1,5) -- (1,1) -- (11,1) -- (11,5) -- (8,5) -- (8,7) -- (11,7) -- (11,11) -- (3,11) -- (3,10) -- (10,10) -- (10,8) -- (7,8) -- (7,4) -- (10,4) -- (10,2) -- (2,2) -- (2,4) -- (5,4) -- (5,8) -- (2,8) -- (2,13)
        ;

\end{tikzpicture}
    \caption{Completed instance example}
    \end{subfigure}
    \hfill
    \begin{subfigure}[t]{0.45\textwidth}
    \centering
    \begin{tikzpicture}[scale=0.44]
    \draw[step=1cm,gray,very thin] (-1,-1) grid (13,13);
    \draw[line width=1pt] (-1,-1) rectangle (13,13);

    \def\blackcells{1/1, 1/4, 1/7, 1/8, 1/10, 2/1, 2/4, 3/1, 3/4, 3/7, 4/1, 4/4, 
4/7, 4/10, 5/10, 6/1, 7/1, 7/4, 7/5, 7/6, 7/7, 7/10, 8/10, 9/4, 9/10, 10/1, 10/2, 10/4, 10/7, 10/8, 10/10, 3/10, -1/-1, -1/0, -1/1, -1/2, -1/3, -1/4, -1/5, -1/6, -1/7, -1/8, -1/9, -1/10, -1/11, -1/12, 0/-1, 1/-1, 2/-1, 3/-1, 3/12, 4/-1, 4/12, 5/-1, 5/12, 6/-1, 6/12, 7/-1, 7/12, 8/-1, 8/12, 9/-1, 9/12, 10/-1, 10/12, 11/-1, 11/12, 12/-1, 12/0, 12/1, 12/2, 12/3, 12/4, 12/5, 12/6, 12/7, 12/8, 12/9, 12/10, 12/11, 12/12}
    \def\redcells{0/0, 0/1, 0/2, 0/3, 0/4, 0/5, 0/6, 0/7, 0/8, 0/9, 0/10, 0/11, 1/0, 2/0, 2/11, 3/0, 3/11, 4/0, 4/11, 5/0, 5/11, 6/0, 6/11, 7/0, 7/11, 8/0, 8/11, 9/0, 9/11, 10/0, 10/11, 11/0, 11/1, 11/2, 11/3, 11/4, 11/5, 11/6, 11/7, 11/8, 11/9, 11/10, 11/11, 1/5, 2/2, 2/3, 2/5, 2/6, 2/8, 2/9, 3/2, 3/3, 3/5, 3/6, 3/8, 3/9, 4/2, 4/3, 4/8, 5/2, 5/3, 5/5, 5/6, 5/8, 5/9, 6/2, 6/3, 6/5, 6/6, 6/7, 
6/8, 6/9, 7/9, 8/2, 8/3, 8/5, 8/6, 8/8, 8/9, 9/2, 9/3, 9/5, 9/6, 9/8, 9/9, 10/5, 6/4, 2/10, 0/12, 2/12}
    \def\blacklines{0/0, 0/4, 0/6, 0/7, 1/0, 1/1, 1/3, 1/4, 1/7, 3/0, 
3/1, 3/3, 3/4, 3/6, 3/7, 3/10, 4/1, 4/3, 4/7, 4/9, 4/10, 6/0, 6/1, 6/3, 6/4, 6/6, 6/7, 6/9, 7/0, 7/4, 7/6, 7/9, 7/10, 9/1, 9/3, 9/4, 9/7, 9/9, 9/10, 10/0, 10/1, 10/4, 10/6, 10/7, 10/10, 2/10, 2/9, -1/-1, -1/0, -1/1, -1/2, -1/3, -1/4, -1/5, -1/6, -1/7, -1/8, -1/9, -1/10, -1/11, 0/-1, 1/-1, 2/-1, 2/11, 3/-1, 3/11, 4/-1, 4/11, 5/-1, 5/11, 6/-1, 6/11, 7/-1, 7/11, 8/-1, 8/11, 9/-1, 9/11, 10/-1, 10/11, 11/-1, 11/0, 11/1, 11/2, 11/3, 11/4, 11/5, 11/6, 11/7, 11/8, 11/9, 11/10, 11/11}

    \foreach \x/\y in \blackcells {
        \fill[black] (\x+0.5,\y+0.5) circle (0.15cm);
    }

    \foreach \x/\y in \redcells {
        \fill[red] (\x+0.5,\y+0.5) circle (0.15cm);
    }

    \foreach \x/\y in \blacklines {
        \fill[black] (\x+1,\y+1) circle (0.15cm);
    }

    \draw[black, ultra thick, dashed]
        (3,0) -- (3,12)
        (6,0) -- (6,12)
        (9,0) -- (9,12)
        (0,3) -- (12,3)
        (0,6) -- (12,6)
        (0,9) -- (12,9)
        ;

    \draw[blue, ultra thick]
        (0,13) -- (0,0) -- (12,0) -- (12,12) -- (3,12) -- (3,13)
        (1,13) -- (1,7) -- (4,7) -- (4,5) -- (1,5) -- (1,1) -- (11,1) -- (11,5) -- (8,5) -- (8,7) -- (11,7) -- (11,11) -- (3,11) -- (3,10) -- (10,10) -- (10,8) -- (7,8) -- (7,4) -- (10,4) -- (10,2) -- (2,2) -- (2,4) -- (5,4) -- (5,8) -- (2,8) -- (2,13)
        ;

        \foreach \x/\y in {1/1, 1/2, 1/3, 1/4, 1/7, 1/8, 1/9, 1/10, 1/11, 1/12, 2/1, 2/4, 2/7, 3/1, 
3/4, 3/7, 3/10, 4/1, 4/4, 4/5, 4/6, 4/7, 4/10, 5/1, 5/10, 6/1, 6/10, 7/1, 
7/4, 7/5, 7/6, 7/7, 7/10, 8/1, 8/4, 8/7, 8/10, 9/1, 9/4, 9/7, 9/10, 
10/1, 10/2, 10/3, 10/4, 10/7, 10/8, 10/9, 10/10} {
        \fill[blue, opacity=0.2] (\x,\y) rectangle ++(1,1);
    }

\end{tikzpicture}
    \caption{Completed instance example (Shaded)}
    \end{subfigure}
    \caption{Example reduction from T-metacell grid (continued)}
		\label{18}
\end{figure}

In the case where the solution consists of type-c as well, the next step is to see that one type-c traversal implies three more adjacent ones since they have to connect by one line. In fact, each type-c cell is a part of exactly one of such block. There are only two possibilities for each corner of the block, making for 16 variants, 2 of which are shown in \cref{19}.

\begin{figure}[H]
    \centering
    \begin{subfigure}[t]{0.45\textwidth}
    \centering
       \begin{tikzpicture}[scale=0.8]
    \draw[step=1cm,gray,very thin] (0,0) grid (6,6);
    \draw[line width=1pt] (0,0) rectangle (6,6);

    \def\blackcells{0/1, 0/4, 1/1, 1/4, 4/1, 4/4, 5/1, 5/4}
    \def\redcells{0/0, 0/2, 0/3, 0/5, 1/0, 1/5, 2/0, 2/2, 2/3, 2/5, 3/0, 3/2, 3/3, 3/5, 4/0, 4/5, 5/0, 5/2, 5/3, 5/5}
    \def\blacklines{0/0, 0/1, 0/3, 0/4, 1/0, 1/4, 3/0, 3/4, 4/0, 4/1, 4/3, 4/4} 

    \foreach \x/\y in \blackcells {
        \fill[black] (\x+0.5,\y+0.5) circle (0.15cm);
    }

    \foreach \x/\y in \redcells {
        \fill[red] (\x+0.5,\y+0.5) circle (0.15cm);
    }

    \foreach \x/\y in \blacklines {
        \fill[black] (\x+1,\y+1) circle (0.15cm);
    }

    \draw[black, ultra thick, dashed]
        (3,0) -- (3,6)
        (0,3) -- (6,3)
        ;

    \draw[blue, ultra thick]
        (0,5) -- (2,5) -- (2,1) -- (0,1)
        (6,5) -- (4,5) -- (4,1) -- (6,1)
        (0,4) -- (6,4)
        (0,2) -- (6,2)
        ;

\end{tikzpicture}
    \caption{A type-c traversal variant}
    \end{subfigure}
    \hfill
    \begin{subfigure}[t]{0.45\textwidth}
    \centering
    \begin{tikzpicture}[scale=0.8]
    \draw[step=1cm,gray,very thin] (0,0) grid (6,6);
    \draw[line width=1pt] (0,0) rectangle (6,6);

    \def\blackcells{0/1, 1/1, 1/4, 1/5, 4/0, 4/1, 4/4, 5/1, 5/4}
    \def\redcells{0/0, 0/2, 0/3, 0/4, 0/5, 1/0, 2/0, 2/2, 2/3, 2/5, 3/0, 3/2, 3/3, 3/5, 4/5, 5/0, 5/1, 5/2, 5/3, 5/5}
    \def\blacklines{0/0, 0/1, 0/3, 0/4, 1/0, 1/4, 3/0, 3/4, 4/0, 4/1, 4/3, 4/4} 

    \foreach \x/\y in \blackcells {
        \fill[black] (\x+0.5,\y+0.5) circle (0.15cm);
    }

    \foreach \x/\y in \redcells {
        \fill[red] (\x+0.5,\y+0.5) circle (0.15cm);
    }

    \foreach \x/\y in \blacklines {
        \fill[black] (\x+1,\y+1) circle (0.15cm);
    }

    \draw[black, ultra thick, dashed]
        (3,0) -- (3,6)
        (0,3) -- (6,3)
        ;

    \draw[blue, ultra thick]
        (1,6) -- (1,4) -- (6,4)
        (0,1) -- (2,1) -- (2,6)
        (6,5) -- (4,5) -- (4,0)
        (5,0) -- (5,2) -- (0,2)
        ;

\end{tikzpicture}
    \caption{A type-c traversal variant}
    \end{subfigure}
    \caption{Two of sixteen possible combinations of type-c traversal}
		\label{19}
\end{figure}

Given a solution to Yagit that includes type-c traversal, we can construct a Hamiltonian cycle. Denote any continuous path of length at least 1 that consists of only type-a or type-b traversal as a section. By definition, the solution is broken down into multiple sections, disrupted by type-c cells. Define completely separate as requiring additional line/s. Since everything is made using one line, it can be assumed that there is no part completely separate from the rest. A cycle consisting only of type-a and type-b would therefore be impossible. A section that connects to a type-c also cannot be a cycle because every other exits of that type-c cell is connected to other type-c cells. We can then assume that each section is a path and has exactly two end points, where an end point can only be one of the following: the entrance, the turning point, or a type-c cell.

The next step is to combine these sections. By definition, each section must have exactly two ending points and the total number of ending points is equivalent to the number of type-c cells plus the entrance and the turning point. For each group of type c-cells, replace the cells with either type-a or type-b to connect the lines horizontally or vertically as shown in \cref{20}.

\begin{figure}[H]
    \centering
    \begin{subfigure}[t]{0.45\textwidth}
    \centering
        \begin{tikzpicture}[scale=0.8]
    \draw[step=1cm,gray,very thin] (0,0) grid (6,6);
    \draw[line width=1pt] (0,0) rectangle (6,6);

    \def\blackcells{0/1, 1/1, 1/4, 1/5, 4/0, 4/1, 4/4, 5/1, 5/4}
    \def\redcells{0/0, 0/2, 0/3, 0/4, 0/5, 1/0, 2/0, 2/2, 2/3, 2/5, 3/0, 3/2, 3/3, 3/5, 4/5, 5/0, 5/1, 5/2, 5/3, 5/5}
    \def\blacklines{0/0, 0/1, 0/3, 0/4, 1/0, 1/4, 3/0, 3/4, 4/0, 4/1, 4/3, 4/4} 

    \foreach \x/\y in \blackcells {
        \fill[black] (\x+0.5,\y+0.5) circle (0.15cm);
    }

    \foreach \x/\y in \redcells {
        \fill[red] (\x+0.5,\y+0.5) circle (0.15cm);
    }

    \foreach \x/\y in \blacklines {
        \fill[black] (\x+1,\y+1) circle (0.15cm);
    }

    \draw[black, ultra thick, dashed]
        (3,0) -- (3,6)
        (0,3) -- (6,3)
        ;

    \draw[blue, ultra thick]
        (1,6) -- (1,4) -- (6,4)
        (0,1) -- (2,1) -- (2,6)
        (6,5) -- (4,5) -- (4,0)
        (5,0) -- (5,2) -- (0,2)
        ;

\end{tikzpicture}
    \caption{Original type-c cells}
    \end{subfigure}
    \hfill
    \begin{subfigure}[t]{0.45\textwidth}
    \centering
        \begin{tikzpicture}[scale=0.8]
    \draw[step=1cm,gray,very thin] (0,0) grid (6,6);
    \draw[line width=1pt] (0,0) rectangle (6,6);

    \def\blackcells{0/1, 1/1, 1/4, 1/5, 4/0, 4/1, 4/4, 5/1, 5/4}
    \def\redcells{0/0, 0/2, 0/3, 0/4, 0/5, 1/0, 2/0, 2/2, 2/3, 2/5, 3/0, 3/2, 3/3, 3/5, 4/5, 5/0, 5/1, 5/2, 5/3, 5/5}
    \def\blacklines{0/0, 0/1, 0/3, 0/4, 1/0, 1/4, 3/0, 3/4, 4/0, 4/1, 4/3, 4/4} 

    \foreach \x/\y in \blackcells {
        \fill[black] (\x+0.5,\y+0.5) circle (0.15cm);
    }

    \foreach \x/\y in \redcells {
        \fill[red] (\x+0.5,\y+0.5) circle (0.15cm);
    }

    \foreach \x/\y in \blacklines {
        \fill[black] (\x+1,\y+1) circle (0.15cm);
    }

    \draw[black, ultra thick, dashed]
        (3,0) -- (3,6)
        (0,3) -- (6,3)
        ;

    \draw[blue, ultra thick]
        (1,6) -- (1,4) -- (6,4)
        (2,6) -- (2,5) -- (6,5)
        (0,1) -- (4,1) -- (4,0)
        (5,0) -- (5,2) -- (0,2)
        ;

\end{tikzpicture}
    \caption{Horizontal connections}
    \end{subfigure}

    \vspace{1em}

    \begin{subfigure}[t]{0.45\textwidth}
    \centering
        \begin{tikzpicture}[scale=0.8]
    \draw[step=1cm,gray,very thin] (0,0) grid (6,6);
    \draw[line width=1pt] (0,0) rectangle (6,6);

    \def\blackcells{0/1, 1/1, 1/4, 1/5, 4/0, 4/1, 4/4, 5/1, 5/4}
    \def\redcells{0/0, 0/2, 0/3, 0/4, 0/5, 1/0, 2/0, 2/2, 2/3, 2/5, 3/0, 3/2, 3/3, 3/5, 4/5, 5/0, 5/1, 5/2, 5/3, 5/5}
    \def\blacklines{0/0, 0/1, 0/3, 0/4, 1/0, 1/4, 3/0, 3/4, 4/0, 4/1, 4/3, 4/4} 

    \foreach \x/\y in \blackcells {
        \fill[black] (\x+0.5,\y+0.5) circle (0.15cm);
    }

    \foreach \x/\y in \redcells {
        \fill[red] (\x+0.5,\y+0.5) circle (0.15cm);
    }

    \foreach \x/\y in \blacklines {
        \fill[black] (\x+1,\y+1) circle (0.15cm);
    }

    \draw[black, ultra thick, dashed]
        (3,0) -- (3,6)
        (0,3) -- (6,3)
        ;

    \draw[blue, ultra thick]
        (1,6) -- (1,2) -- (0,2)
        (0,1) -- (2,1) -- (2,6)
        (6,5) -- (4,5) -- (4,0)
        (6,4) -- (5,4) -- (5,0)
        ;

\end{tikzpicture}
    \caption{Vertical connections}
    \end{subfigure}
    \caption{Replacement method of one variant}
		\label{20}
\end{figure}

Every replacement removes one block of exactly four type-c cells. Since every type-c cell belongs to exactly one of such structure, the final result has no type-c cells, and must only have two end points: the entrance and the turning point. After each replacement, two sections is either combined into a longer path, or a cycle is formed. The path connected to the entrance or the turning point cannot be in a cycle, which means the sections must now include a path from the entrance to the turning point and possibly $O(n)$ separate cycles each of length at least four. This is true for any arbitrary order and replacement choices. 

Since there were no cycles consisting of just type-a and type-b cells prior to the replacement, the cycles that currently exists must include cells that were previously type-c. Notice that any of such cycles must contain two of the four type-c cells for each block replacement. Furthermore, they must be connected to the other two ex type-c cells from that block which are now part of another section. By replacing each such cells from type-a to type-b or vice versa, the two sections merge into one continuous section. Merging two cycle will form a cycle, and merging the path to a cycle will lengthen the path. 

Denote this ability to connect two sections as the two being connectable, the initial number of cycles as $N$, and the section connecting the entrance to the turning point as the main section. Next, we will prove the existence of the Hamiltonian cycle using this method of replacement. 

\begin{proof}
We will prove by induction on the number of replacements.

\textbf{Base Case:} Before any replacement, no modification has been made, and every property is satisfied as explained earlier.

\textbf{Inductive Step:} Assume that after $k$ replacements, $N - k$ cycles remain in the solution, and that either the main section traverses every T-metacells, or there is at least one pair of connectable sections.

Consider the $(k+1)$-th replacement. By assumption if the main section doesn't traverse every cell, a cycle can be combined with another section, reducing the number of cycles from $N - k$ to $N - (k+1)$. Notice that a block of four type-c cells can only be a part of two sections, therefore, the cells that were replaced cannot connect the combined section to another section and needs no further modification afterward. Therefore, once combined, two cycles will not be separated, and the number of cycles remaining will always decrease with each replacement. 

If the main section traverse every T-metacells, we are done. Otherwise, there exist at least one cycle. This cycle must be connectable to another section, else, that cycle must have been completely separate from the start, a contradiction.

By the inductive hypothesis, this results in a path from the entrance to the turning point that traverses every T-metacells and contains only type-a and type-b cells, thus creating a Hamiltonian cycle in the original graph.
\end{proof}

Thus, Yagit is NP-complete. However, because of the type-c traversal, local solutions of Yagit are not always unique, and this design is not sufficient to show ASP-completeness.

\section{Conclusion}
Through the T-metacell framework, we proved the ASP-completeness of three puzzles, implying their NP-completeness as well. We then proved the NP-completeness of one additional puzzle.

\bibliographystyle{plainurl}
\bibliography{references}
\end{document}